\documentclass[twocolumn, 
               a4paper,
	       floatfix,
               superscriptaddress, 
               showpacs, showkeys,
               rmp, amssymb, amsmath]{revtex4-1}

\usepackage{graphics}
\usepackage{graphicx}
\usepackage{color}
\usepackage[dvips]{epsfig}
\usepackage{amssymb}
\usepackage[normalem]{ulem}

\newcommand{\de}{\mathrm{d}}
\newcommand{\GDRADN}{Nuclear Architecture and Dynamics, CNRS GDR 3536, UPMC Universit\'e Paris 6, 75005 Paris France}
\newcommand{\GDRGSI}{Genomes Structure and Instability, Sorbonne Universit\'es,
National Museum of Natural History, Inserm U 1154, CNRS UMR 7196, 75005 Paris
France}
\newcommand{\UPMC}{Sorbonne Universit\'es UPMC Univ. Paris 06 UMR 7600  LPTMC F-75005 Paris France.}
\newcommand{\LPTMC}{CNRS UMR 7600 LPTMC, F-75005 Paris France}
\newcommand{\IGMM}{Institut de G\'en\'etique Mol\'eculaire de Montpellier CNRS UMR 5535 Montpellier France.}

\begin{document}

\title{The physics of epigenetics}
\date{\today}


\author{Ruggero Cortini} 
\affiliation{\UPMC}
\affiliation{\LPTMC}
\affiliation{\GDRADN}

\author{Maria Barbi}
\affiliation{\UPMC}
\affiliation{\LPTMC}
\affiliation{\GDRADN}

\author{Bertrand R. Car\'e}
\affiliation{\UPMC}
\affiliation{\LPTMC}
\affiliation{\GDRADN}

\author{Christophe Lavelle} 
\affiliation{\GDRADN}
\affiliation{\GDRGSI}

\author{Annick Lesne}
\affiliation{\UPMC}
\affiliation{\LPTMC}
\affiliation{\GDRADN}
\affiliation{\IGMM}

\author{Julien Mozziconacci},
\affiliation{\UPMC}
\affiliation{\LPTMC}
\affiliation{\GDRADN}

\author{Jean-Marc Victor}
\altaffiliation{Corresponding author: Jean-Marc Victor LPTMC case courrier 121 Universit\'e Pierre et marie Curie 4 place Jussieu 75252, Paris cedex 05 France. Email: victor@lptmc.jussieu.fr}
\affiliation{\UPMC}
\affiliation{\LPTMC}
\affiliation{\GDRADN}
\affiliation{\IGMM}


\begin{abstract}
In higher organisms, all cells share the same genome, but every cell expresses
only a limited and specific set of genes that defines the cell type.  During cell
division, not only the genome, but also the cell type is inherited by the
daughter cells. This intriguing phenomenon is achieved by a variety of processes
that have been collectively termed epigenetics: the stable and inheritable
changes in gene expression patterns. This article reviews the extremely rich and
exquisitely multi-scale physical mechanisms that govern the biological processes
behind the initiation, spreading and inheritance of epigenetic states. These
include not only the changes in the molecular properties associated with the
chemical modifications of DNA and histone proteins, such as methylation and
acetylation, but also less conventional changes, typically in the the physics
that governs the three-dimensional organization of the genome in cell nuclei.
Strikingly, to achieve stability and heritability of epigenetic states, cells
take advantage of many different physical principles, such as the universal
behavior of polymers and copolymers, the general features of dynamical systems,
and the electrostatic and mechanical properties related to chemical
modifications of DNA and histones. By putting the complex biological literature
under this new light, the emerging picture is that a limited set of general
physical rules play a key role in initiating, shaping and transmitting this
crucial ``epigenetic landscape''.  This new perspective not only allows to
rationalize the normal cellular functions, but also helps to understand the
emergence of pathological states, in which the epigenetic landscape becomes
dysfunctional.
\end{abstract}

\maketitle

\tableofcontents

\section{Introduction}
\label{sec:intro}

\subsection{An intricate history}

\begin{figure*}
 \includegraphics[width=\textwidth]{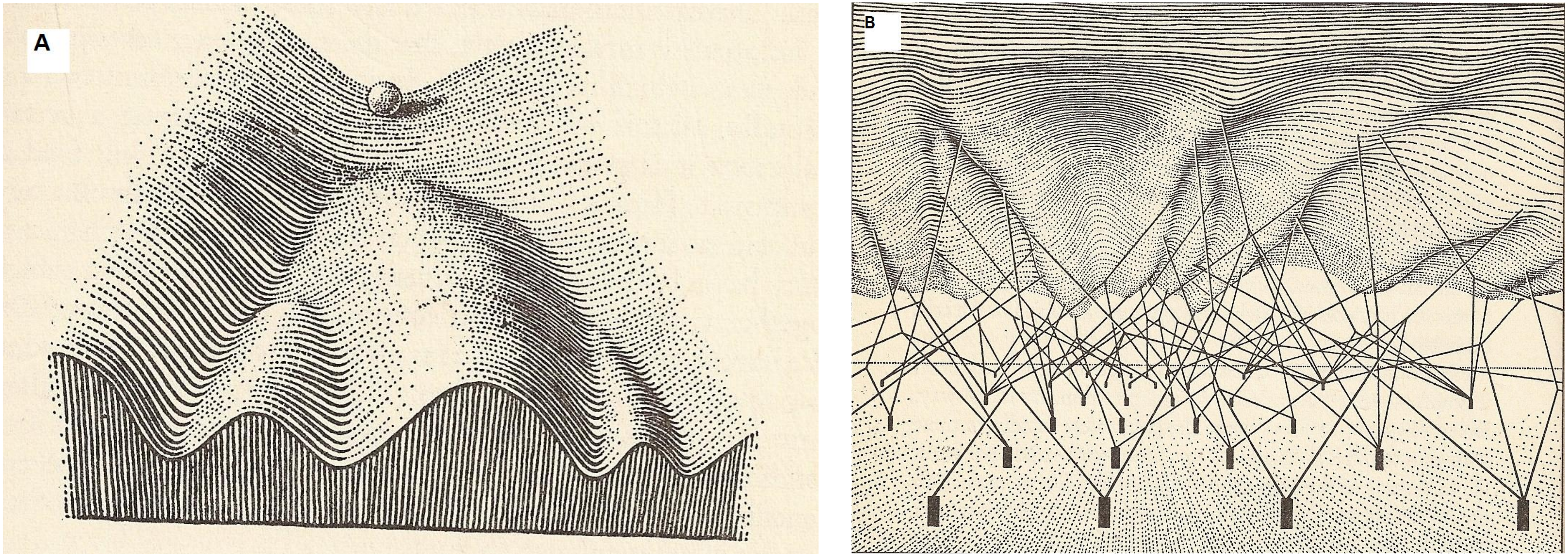}%
  \caption{
    (Left) The epigenetic landscape described by Waddington
    \cite{Waddington1957} represents the process by which the cell (represented
    by the ball) faces different possible paths during development (i.e.  choose
    one of the permitted trajectories), leading to different cell fates. (Right)
    The landscape is dynamically determined by hidden wires that symbolize genes
    expression and interactions. See Sec.~\ref{seq:biological-relevance} and
    Sec.~\ref{sec:conclusion} for a physical reanalysis of this metaphorical
    picture. Adapted from \cite{Waddington1957}.
  \label{fig:Waddington}}
\end{figure*}

The word ``epigenetic'' has been introduced by Waddington in  1942 in the
context of development, to qualify all the processes relating the genotype and
the phenotype of an organism  \cite{Waddington1942}. The associated
investigations belonged to the domain, novel at that time, of developmental
genetics.  The word  ``epigenetic'' in this original meaning  was imprinted by
the pre-existing  concept of epigenesis, namely the fact that the organism is
not fully achieved in the initial cell but experiences complex developmental
processes \cite{Gilbert2011}. Epigenetics was at that time the mechanistic study
of how genes guide  the epigenesis (development) of an organism, what is
captured in a metaphoric way by the famous Waddington's epigenetic landscape: a
landscape, shaped by the genes, on which the organism would evolve during its
development as  a rolling stone on the landscape, following one of the possible
epigenetic pathways (see Fig.~\ref{fig:Waddington}). 

In parallel, the adjective has been used with the meaning of ``para-genetic''.
Epigenetic systems, as opposed to the  genetic system,  were conceived as
``signal interpreting devices'' \cite{Nanney1958}, i.e. mediators between
signals -- environmental or physiological cues -- and the genomic response,
mainly at the level of transcriptional regulation.

Due to this dual origin of the word ``epigenetic'', the associated concepts have
developed in several ways (see \cite{Haig2004} for a detailed historical
account) and one could find in 1994 the two following complementary definitions
of epigenetics \cite{Holliday1994}: (1) changes in gene expression which occur
in organisms with differentiated cells, and the mitotic inheritance of the
associated patterns of gene expression; (2) transgenerational inheritance, that
is, transmission through meiosis of non-genomic information.

Due to this intricate history, a consensus definition of epigenetics is still
lacking today \cite{Dawson2012}. Notably the transgenerational inheritance,
albeit largely documented in plants, remains a matter of debate in animals and
especially in humans.

Recently, some authors proposed an operational definition of epigenetics: ``An
epigenetic trait is a stably heritable phenotype resulting from changes in a
chromosome without alterations in the DNA sequence''.  This will be the
definition we use in this review. To be even more specific, we note that:

Epigenetics is the modification of the function(s) of a gene, that is stable
and heritable during mitosis, possibly during meiosis.

Epigenetics is not the reversible regulation of transcription in response to
metabolic cues, because this is \emph{not stable} nor \emph{heritable}.

\subsection{Scope of this review} 
 
In this review, we not only intend to analyze the physics that drives or
accompanies epigenetic marking, but we also aim at understanding the
rationale behind this marking. And physics is a beautiful, yet underrated guide
to reach this goal.

Several epigenetic mechanisms will be distinguished: those occurring  at the
level of DNA, those involving histone post-translational modification, and less
conventional ones involving  chromatin topology (supercoiling) and nuclear
architecture.

We first introduce in section \ref{sec:chromatin} the physical template of
epigenetic marking, namely chromatin.

Section \ref{sec:regul_BY} is devoted to the physics behind the family of
processes at work in the way epigenetic marks control gene expression in
different cell types. 

Section \ref{sec:regul-OF} addresses the issue of the initiation, spreading,
maintenance, and heritability of the epigenetic marks in the framework of
dynamical systems.

In Section \ref{sec:more} we review other epigenetic processes that have a
less clear-cut physical interpretation: DNA methylation, imprinting, chromosome
X inactivation, supercoiling marking. In conclusion we finally propose a list of
currently significant and challenging issues.

Due the fundamentally different logic of transcriptional regulation in
prokaryotes and eukaryotes \cite{Struhl1999}, we will let aside the realm of
bacteria, although epigenetic switches have been observed as well in prokaryotic
cells and have been modeled successfully \cite{Lim2007,Norregaard2013}.

We hope this review will be a stimulating introduction to epigenetics for
physicists as well as an ``alternative reading frame'' of epigenetics for
biologists that will help tackling cutting-edge advances in current topics
ranging from nuclear organization and cell differentiation up to cancer
progression and chronic diseases.

\section{The physical template of epigenetics: chromatin}
\label{sec:chromatin}
In all living organisms, DNA encodes the genetic instructions required to
synthesize proteins, the basic bricks ensuring the proper functioning of
the cell. The main steps of protein synthesis are DNA transcription into RNA,
then RNA translation into an aminoacid chain and chain folding to form a
functional protein \cite{Alberts2013}.

The very same genome is found in each cell. It has to be packaged inside
its tiny volume, and has to be retrieved at will for physiological purposes. DNA is
therefore embedded in an orderly and dynamically retrievable architecture. Two
main organizational strategies can be identified. In prokaryotes (bacteria), DNA
is located in the same compartment as all other intracellular components.  In
eukaryotes (from the unicellular yeast up to multicellular organisms, including
fungi, animals and plants), DNA is sequestered in the nucleus, a dedicated
compartment enclosed within a membrane.

In the cell nucleus, multiple long linear DNA molecules are organized by
architectural proteins to form chromosomes. From a physicist point of view,
chromosomes are giant polymers.  During mitosis, i.e. cell division, chromosomes
duplicate and then condense in the well known ``X'' shape, with each DNA copy
forming one of the two rods (the sister chromatids, bound together at the
centromere). The rest of the time (i.e. during interphase), chromosomes are less
condensed and fill the whole nucleus, more or less homogeneously
\cite{Leblond1998}. To give a quantitative idea of the composition of an
interphase nucleus, the dry matter of a yeast nucleus is about $\sim$70-80\% in
protein, $\sim$20-30\% in RNA, and only $\sim$2\% in DNA \cite{Rozijn1964}.

In this section we introduce the basic concepts that come into play in the study
of epigenetics. In Sec.~\ref{subsec:nucleosome} we give an overview of the
molecular structure of chromatin, and we introduce the concept of epigenetic
marks. In Sec.~\ref{subsec:chromatin_largescale} we will give an overview of
the large-scale organization of chromatin in the cellular nucleus, stressing
the importance of this organization in gene expression. Finally, in
Sec.~\ref{subsec:chromatin_as_polymer} we give a synthetic picture of these two
aspects, in the framework of polymer physics.

\subsection{Molecular picture of chromatin and its modifications}
\label{subsec:nucleosome}

\begin{figure*}
  \includegraphics[width=\textwidth]{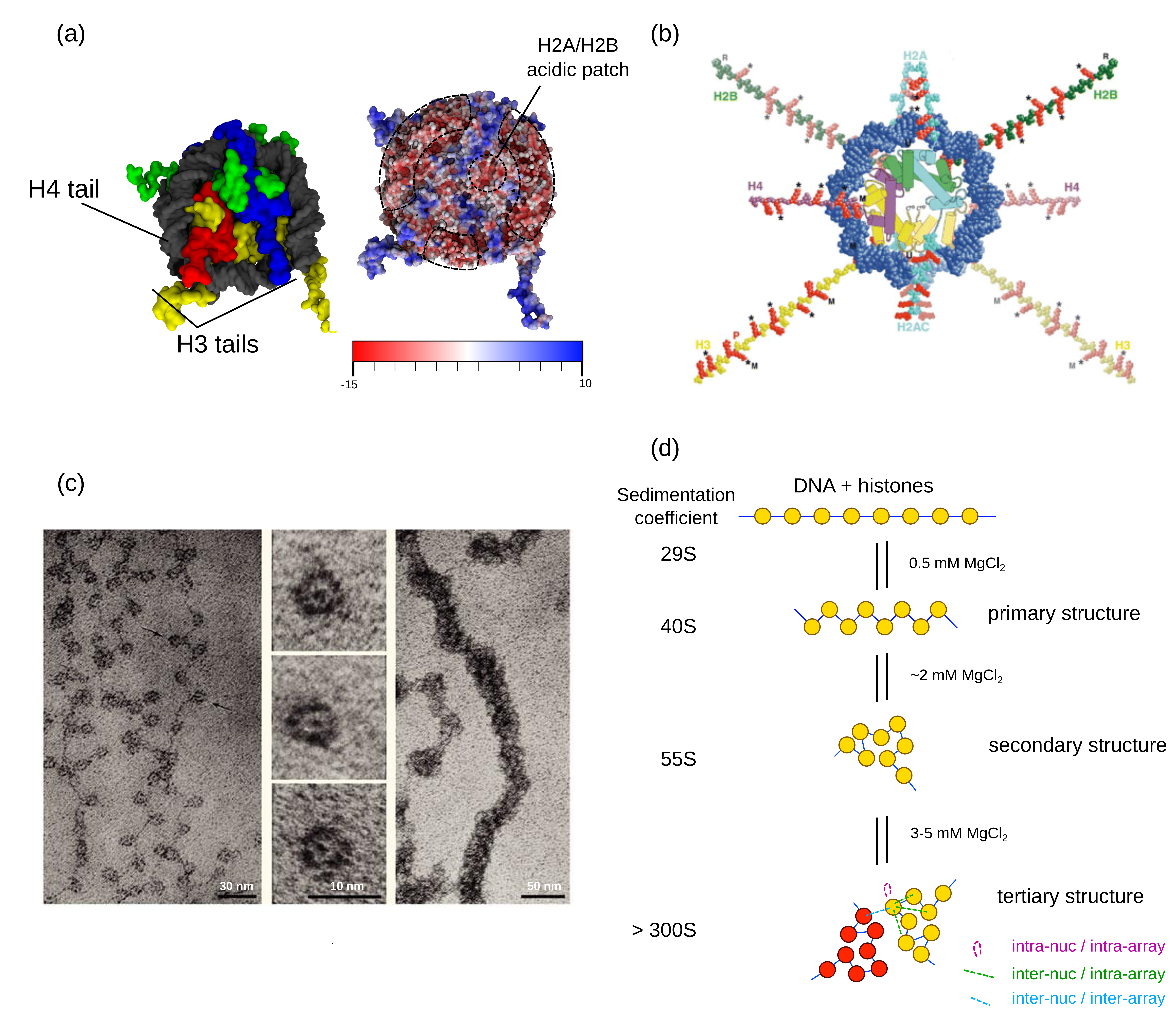}
  \caption{Detailed structure of the Nucleosome Core Particle (NCP) and chromatin
    fiber. (a) left: internal tertiary structure of the NCP147 (PDB: 1KX5) H2A
    is in green, H2B in blue, H3 in yellow, H4 in red, and DNA in black. right:
    electrostratic potential at the NCP surface (computed using the PDB2PQR/APBS
    plugin \cite{Dolinsky2007, Baker2001} of the Visual Molecular Dynamics
    (www.ks.uiuc.edu/Research/vmd) software package\cite{Humphrey1996},
    ionic strength 0.15 M monovalent salt). (b) Cartoon of the NCP showing
    histone tails and the globular core. Lysine and arginines residues are
    marked by asterisks (from \cite{Wolffe1999}). (c) Electron Microscopy images
    of nucleosome arrays with low (left) or high (far right) ionic forces, with
    details of nucleosome (middle), from \cite{Olins2003}. (d) Reconstituted
    chromatin sedimentation assays principles, adapted from
\cite{Pepenella2013}.\label{fig:nuc_details}}
\end{figure*}

In eukaryotic organisms, chromosomal DNA is associated with proteins to form
chromatin. The principal proteins associated with DNA are called histones.
Histones are polypeptidic monomers of five types: Histone 1 (H1) class, Histone
2A (H2A) class, Histone 2B (H2B) class, Histone 3 (H3) class and Histone 4 (H4)
class. Each histone family has variants whose presence in chromatin depends on
the species, the cell type, and the development stage.  The classical structure
of the histone-DNA assembly consists of 1.7 left-handed turns of double strand
DNA (approximately 147 base pairs or bp) wrapped around a histone octamer
composed of two copies of each histone monomer H2A, H2B, H3 and H4
\cite{Luger1997,Davey2002}. In most species, this assembly, referred to as the
nucleosome core particle (NCP) (see Fig. \ref{fig:nuc_details}a,b),  may also
integrate a copy of H1 (linker histone) at
the DNA entry/exit point, although H1 does not share the ubiquity of the other
histone classes. 

In addition, consecutive NCPs are separated by linker DNA whose length ranges
from 20 to 60 bp. Indeed, chromosomes are a succession of NCPs and DNA linkers.
The basic structural unit (monomer) is made of one NCP and one DNA linker, and
is called the nucleosome.  The number of DNA base pairs inside one nucleosome
is the Nucleosome Repeat Length (NRL) (see Fig.~\ref{fig:nuc_details}c,d),
which is not constant and may vary along the genome and across various tissues.

Electrostatic interactions are important because the NCP has a charge of
-150{\it e}, to which DNA contributes -294{\it e} and histones +144{\it e}. The
NCP is therefore not electrically neutral, so the folding of nucleosome arrays
is highly dependent on the presence of positive counterions
\cite{Yang2011,Bertin2007a}. Additionally, the charge distribution in the NCP is
not spatially homogeneous (see Fig.~\ref{fig:nuc_details}a).

Epigenetic marks are chemical covalent modifications of either DNA (namely DNA
methylation, see Sec.~\ref{sec:DNA-methylation}), or histones (so-called
post-translational modifications, PTMs, see Sec.~\ref{sec:histonePTMs}). The DNA
methylation state and the histone PTMs are transmitted through cell division
both because they are covalent and thanks to specific mechanisms. DNA
methylation is accurately transmitted by a specific molecular mechanism (see
Sec.~\ref{sec:DNA-methylation}). Histone PTMs are inherited in a fundamentally
different way, which will be the principal subject of Sec.~\ref{sec:regul-OF}.

\subsection{Large-scale picture of chromatin}
\label{subsec:chromatin_largescale}

Eukaryotic chromosomes are giant polymers, each formed by a huge string of
nucleosomes. The conformation of this string at different length scales is
generally described using an analogy with proteins: the string of nucleosomes
itself can be viewed as the primary structure of chromatin; the conformation
adopted by an array of a few dozen successive nucleosomes forms the secondary
structure of chromatin. The 3D structural arrangement of several arrays can
finally be viewed as the chromatin tertiary structure
\cite{Luger2012,Pepenella2013}, see Fig.~\ref{fig:nuc_details}d.

\begin{figure*}[hp]
  \includegraphics[width=\textwidth]{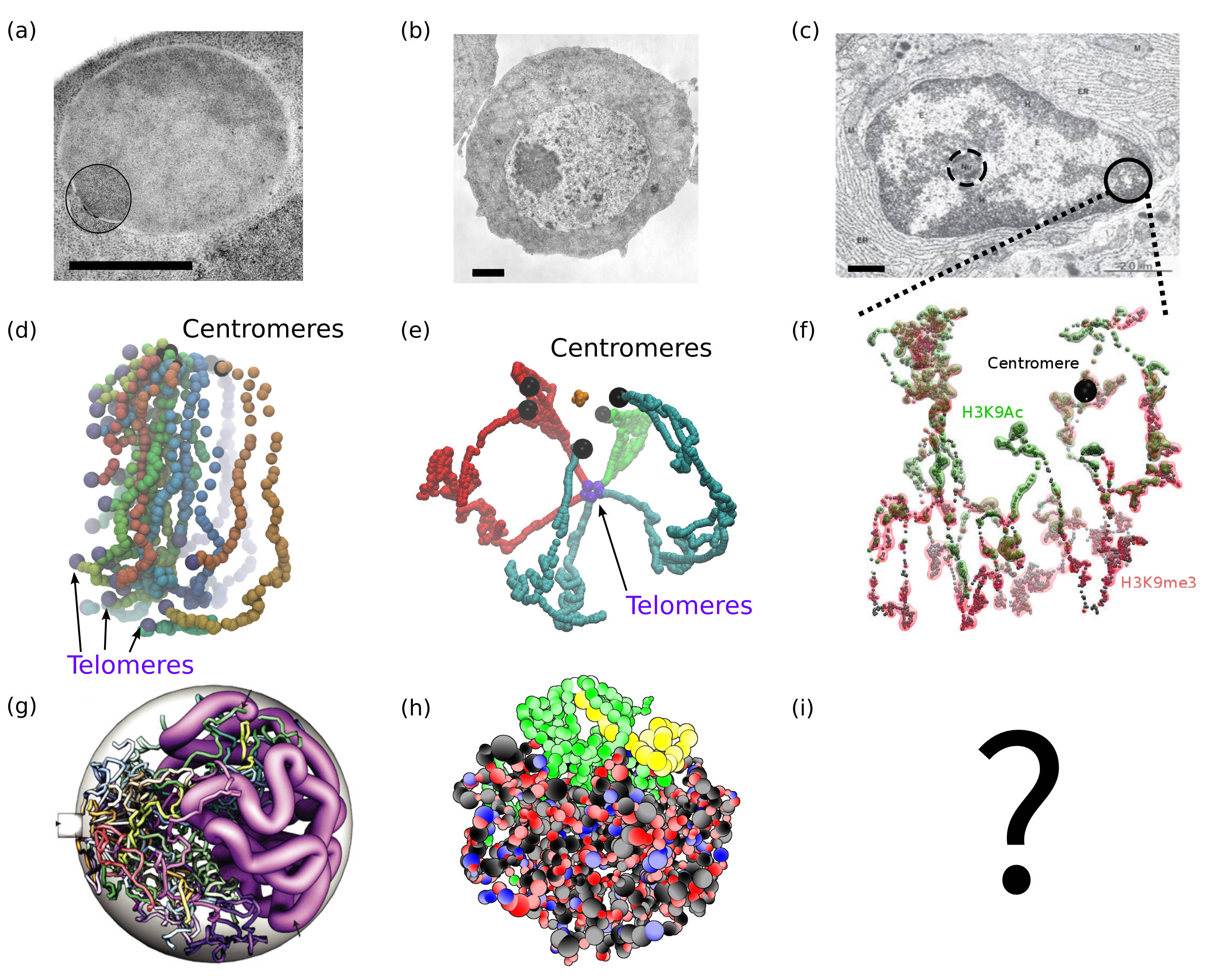}
  \caption{
    Nuclear organization in yeast (\textit{S. Cerevisiae}), drosophila and
    mammals.  (a-c) Electron microscopy images of nuclei in yeast (a) in
    drosophila (b) and human (c). Scale bars correspond respectively to 1
    $\mu$m, 2 $\mu$m and 2 $\mu$m. In (a), the nucleolus is the darker region in
    the upper part of the nucleus. The SPB is shown by a circle. In (b), the
    nucleolus is the dark circular region and heterochromatin can be seen as
    darker spots. In (c), the nucleolus is marked by a dashed circle. (Figure
    (a) from  {\tt
    http://scienceblogs.com/transcript/2006/08/16/the-centrosome-and-the-spindle/},
    (b) from {\tt http://pixgood.com/nuclear-pore-em.html}, (c) from {\tt
    http://tinyurl.com/m6phpf8}).
    (d-f) 3D models reconstruction \cite{Lesne2014} from contact maps obtained
    using the Hi-C protocol in these three organisms \cite{Duan2010, Sexton2012,
    Dixon2012}.  In (d) and (e) all the chromosomes are represented with
    different colors.  In (f), only chromosome 1 is shown and the colors
    correspond to regions harboring different epigenetic marks: H3K9Ac and
    H3K9me3 (see Sec.~\ref{sec:histonePTMs}).  On each reconstruction,
    centromeres are shown as black beads and telomeres (chromosome ends) as
    purple beads.  Each bead represents respectively 12 kb, 40 kb and 40 kb.
    (g-h) Polymer models of the genome in yeast and drosophila: In (g) each
    chromosome is labeled with a different color (from \cite{Wong2012}).  In 
    (h) colors correspond to the colors of chromatin (see
    Sec.~\ref{sec:epigenome}, courtesy of Giacomo Cavalli and Pascal Carrivain
    (IGH, Montpellier)).  (i) To our knowledge, physical models of the human
    genome have not been developed so far.
  \label{fig:nucleus}}
\end{figure*}

When observed by electron microscopy (see Fig.~\ref{fig:nucleus}a,b,c),
interphase chromatin appears to fill the entire nucleus volume. As genome length
may vary considerably from organism to organism, the nucleus size varies
accordingly: orders of magnitude go from $\sim 10 ~\mathrm{Mb}$ (Mega base
pairs) for a diameter of the order of 2 $\mu$m in yeast
(Fig.~\ref{fig:nucleus}a), to $\sim 100 ~\mathrm{Mb}$ and 4 $\mu$m  in
drosophila fly (Fig.~\ref{fig:nucleus}b), and up to $\sim 1000 ~\mathrm{Mb}$
and 10 $\mu$m in mammals (Fig.~\ref{fig:nucleus}c). These differences in size
are certainly correlated with the differences in chromatin organization that
can be directly deduced by simple inspection of electron microscopy images.

Yeast nuclei are the most homogeneously filled ones, with a large, denser region
called the nucleolus, which is known to be the site of very intense
ribosomal RNA synthesis. A smaller, dark linear body can also be seen in the
inset, connected with a star-shaped structure, the spindle pole body (SPB) from
which tubular proteic assemblies, microtubules, stem and ``hold'' chromosomes at
their centromeres. In contrast with multicellular organisms, in yeast this
microtubule bundle is preserved all along the cell cycle. It is a crucial
organization center for the assembly of chromosomes in interphase and for
chromosome segregation during mitosis (see Fig.~\ref{fig:nucleus}a,d,g).

\begin{figure}
  \includegraphics[width=0.5\textwidth]{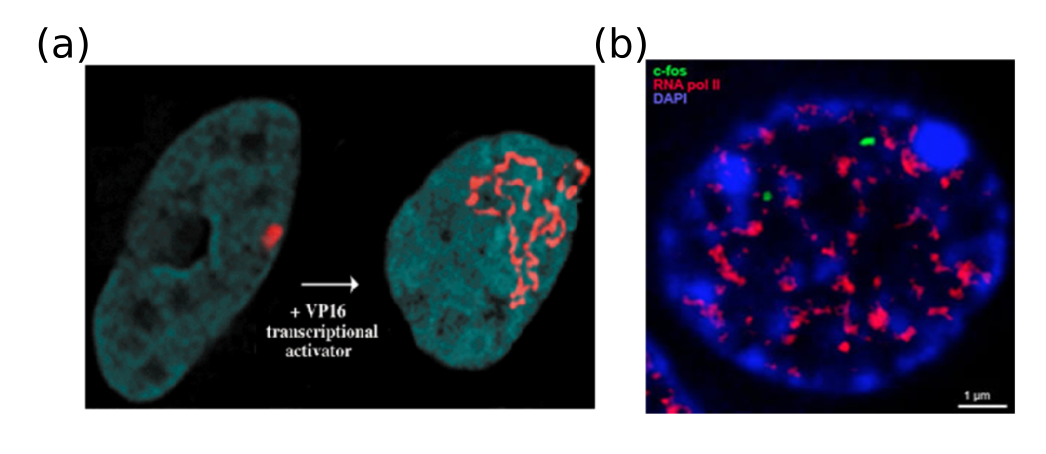}
  \caption{Fluorescence microscopy of nuclei. (a) a specfic region of the
  genome, labeled in red, is decompacted after the induction of transcription.
  (adapted from \cite{Tumbar1999}) (b) A human cell nucleus. DNA is labeled in
  blue, PolII in red (from \cite{Crepaldi2013})  \label{fig:transcription}}
\end{figure}

When the nuclei of multicellular organisms are considered
(Fig.~\ref{fig:nucleus}b,c), their most striking feature is the coexistence of
distinct denser and less compact regions. These regions are persistent and are
not simply the result of temporal fluctuations of chromatin density.  These
features have been shown to strongly correlate with the transcription activity
of genes.  Active genes tend indeed to gather at the center of the nucleus, in a
region where chromatin is less dense and more accessible, which is called
euchromatin. Inactive genes are found instead in denser regions, called
heterochromatin, and tend to associate with the nuclear periphery. As a stunning
example of chromatin compaction and localization changes induced by
transcription, the activation of a genomic locus results in a dramatic change of
its topology (Fig.~\ref{fig:transcription}a).

With the improvement of imaging and labeling techniques, gene transcription by
the RNA polymerase PolII in multicellular organisms has been shown to occur in
well-defined loci, called factories (Fig.~\ref{fig:transcription}b)
\cite{Jackson1993}. These factories are located within the euchromatin domain
and each factory has a propensity to gather co-regulated genes
\cite{Jackson1998}. In this picture, it appears that the functional
differences between cell types are related to the way the genome is folded in
the nucleus of these cells.

In the last two decades, impressive advances in experimental techniques in
measurements of 3D chromosomal contacts have been made, starting from the
``Chromosome conformation capture'' approach \cite{Dekker2002}. Its genome wide
derivative (Hi-C) enables the generation of contact maps at the genome scale
\cite{Lieberman2009}. From these maps, it is possible to reconstruct the
underlying 3D structure of the genome and such structures are represented on
Fig.~\ref{fig:nucleus}d,e,f. The results confirmed the tethering of
centromeres in yeast and drosophila, but also of telomeres. In humans the
reconstruction of chromosome 1 gives a visual illustration of
decondensed euchromatin loops emanating from globular heterochromatin globules.
These are decorated with two different specific histone marks. We will come back
on the results of these investigations in Sec.~\ref{sec:HiC}.

\subsection{Chromosomes as polymers}
\label{subsec:chromatin_as_polymer}

Most of the modeling efforts addressing the question of the nuclear
organization have been so far oriented by polymer physics. The question then
arises as to understand whether polymer physics is the main player that drives
chromosome organization.

(i) In the simplest case of yeast, where chromosomes are shorter and all
anchored at the SPB by their centromeres, it seems to be the case.
Indeed, several polymer simulations have been able to reproduce the structure of
interphase yeast nuclei \cite{Wong2012,Tjong2012}, see Fig.~\ref{fig:nucleus}g
\footnote{Note that, in yeast, the whole genome is actively transcribing most of
the time, with the only exception of the regions that govern the cell sexual
behavior, called ``hidden mating type loci'', and of the  chromosome
extremities, called telomeres, which protect the ends of the chromosome from
damaging or from fusion with other chromosomes. Therefore heterochromatin is
restricted to telomeres and mating type loci in this case.}.  Moreover,
fluorescent microscopy has been used to check the dynamical behavior {\it in
vivo} of given chromosomal loci, \cite{Hajjoul2013,Albert2013}. Single
particle tracking has revealed a quite uniform response within the genome,
characteristic of polymers in confined spaces. Except for telomeres and for
the highly transcribed DNA in the nucleolus, yeast chromosomes behave as a
polymer brush, and are essentially organized by simple physical principles
\cite{Huet2014} (see Fig.~\ref{fig:nucleus}d,g).

(ii) In the well-studied, intermediate-size case of the drosophila, recent
investigations tend to indicate that this polymer behavior is partially
conserved, but with some significant changes that complexify the picture (see
Fig. \ref{fig:nucleus}e,h). Roughly speaking, it has been proposed that
euchromatin and heterochromatin have intrinsically different biochemical and
physical properties, due to a deeply different protein ``dressing'' of the DNA
molecules. More precisely, Filion and co-workers have identified five principal
chromatin states, called chromatin ``colors'' from the analysis of 53 chromatin
protein genome-binding profiles in drosophila cells \cite{Filion2010}. Among
these states, some essentially correspond to active, transcribing euchromatin,
other to dense, repressed heterochromatin.  These chromatin states result from
the recruitment of DNA-binding proteins that are specific of the underlying
epigenetic marks (see Sec.~\ref{sec:histonePTMs}). Note, however, that the
colors are only an arbitrary choice of the authors and do not correspond to
actual coloring of the chromatin, due, e.~g.~, to staining.

As a consequence, drosophila chromosomes are more properly described as
co-polymers, i.e. polymers containing more than one type of monomer. A model of
the resulting copolymer brush is depicted on Fig.~\ref{fig:nucleus}h.

(iii) In mammals heterochromatin is mainly located at the nuclear membrane and
euchromatin at the center of the nucleus (see Fig.~\ref{fig:nucleus}c).
The reconstituted 3D structure of chromosome 1 (the longest human chromosome) shows
an alternance of long loops of euchromatin and dense parts of heterochromatin
tethered to the nuclear membrane (see Fig.~\ref{fig:nucleus}f).

In summary, the conformation adopted by chromatin is affected by its intrinsic
structural parameters such as the NRL (the reader may find an extensive review
in \cite{Boule2015}), on top of which lies an additional layer of modulation by
internucleosomal electrostatic interactions \cite{Hansen2002,Pepenella2013} and
binding of architectural proteins. This conformation is essential for gene
regulation. The epigenetic marks present on DNA and histones, by mediating
specific interactions between portions of chromatin, alter its conformation and
hence its function. The next section will be devoted to understanding the
complex relationship between epigenetic marking and genome structure and
function.

\section{From epigenetic marks to regulation of gene expression through the 3D
organization of the genome}
\label{sec:regul_BY}

\subsection{General principles of gene silencing. The paradigm of DNA accessibility}

During development, the determination of the cell type (cell fate) involves
progressive restrictions in its developmental potency and results from
differential gene expression. DNA methylation is a key control parameter of this
process: genes that are specific for the desired tissue are kept unmethylated,
whereas the others are methylated. Moreover, patterns of DNA
methylation are faithfully propagated throughout successive cell divisions (see
Sec.~\ref{sec:DNA-methylation}). However the physics of DNA methylation
is still elusive and we therefore postpone further developments on DNA
methylation to the last part of this review (see Sec.~\ref{sec:DNA-methylation}).

Epigenetic regulation of gene expression involves \emph{silencing}, i.e. a
permanent and heritable inhibition of gene transcription (transciptional gene
silencing) or translation (post-transcriptional gene silencing). The current
paradigm is that gene silencing is achieved through chromatin condensation, in a
so-called heterochromatinization process \cite{Grewal2003}.  Can we characterize
the physical properties of heterochromatin and euchromatin? What are the
physical consequences of heterochromatinization in terms of structure, dynamics
and how do these physical consequences turn out into functional consequences?

Histones simultaneously play a crucial role in determining the structure of
chromatin; they are the substrate of a vast catalog of epigenetic markings
\cite{Kouzarides2007, Cantone2013}, which is not a coincidence. This supports
the hypothesis that epigenetic histone marks modulate gene expression through
chromatin structural rearrangements at each level of the nuclear organization:
nucleosome, chromatin fiber, chromatin loops, chromosome territories, whole
nucleus \cite{Zhou2007,Poirier2009}.

The potentially huge number of combinations of epigenetic marks
has led to the hypothesis that transcription factors (TFs) might be targeted to
nucleosomes endowed with specific combinations of histone tail
post-translational modifications.  This was coined as the ``histone code''
hypothesis at the beginning of the millenium and remained popular in biology for
the better part of a decade.While the genetic code encodes the sequence of a
protein using the four bases of the DNA, this histone code would encode the
regulatory events involved in triggering its expression, and in particular the
differential expression observed in cells of different types
\cite{Jenuwein2001}. However, it was then realized that only a few combinations
of histone marks are actually observed \cite{Lennartsson2009, Rando2012}. These
relevant combinations can even be reduced to 5 classes in the case of
drosophila, the so called ``five colors of chromatin'' as identified by Filion
et al.~\cite{Filion2010}.  Moreover, it was also realized that the histone code
is actually not a genuine code insofar as the correspondence between the
codewords (combinations of epigenetic marks) and their meaning (the TFs they
code for) is not gratuitous \cite{Lesne2006,Kuhn2014}. Instead, the
correspondence is a biochemical recognition of the pattern of histone marks by
the TF. This is why it was recently proposed that chromatin may have evolved as
an allosteric enzyme able to mediate a gratuitous correspondence between
epigenetic marks and TF binding to the underlying DNA sequence \cite{Lesne2015}.

\subsection{Histone modifications as chromatin structural modulators}
\label{sec:histonePTMs}

Most epigenetic marking occurs on the histones that coat DNA. What are the
physical consequences of this marking and what is its effect on chromatin
organization? 

\subsubsection{Histone tails and their role in internucleosomal interactions}

As already mentioned, nucleosomes are formed by wrapping DNA around an octameric
protein assembly formed by histone proteins.  The N-terminal sequences of H2A,
H3 and H4 extend from the globular histone core to form the so-called histone
tails (see Fig.~\ref{fig:nuc_details}b). The H3 and H4 tails consist
respectively of 35 and 20 residues, of which respectively 13 and 9 are
positively charged (lysines, K and arginines, R). These tails are intrinsically
disordered protein domains, hence adopt a random coil configuration, as
suggested by crystallographic studies \cite{Luger1997, Davey2002} and
proteolytic cleavage assays. Tails contribute differently to intranucleosomal
stability and internucleosomal interactions \cite{Allan1982,Arya2006,
Arya2009,Zhou2007,Sinha2010}. The two H3 tails exit from the histone core close
to the DNA entry-exit site of the nucleosome, and associate preferentially with
DNA to ``lock'' its wrapping around the histone core. The H4 tails are known to
associate with a set of seven residues referred to as the H2A/H2B acidic patch,
located on the H2A-H2B interface (see Fig. \ref{fig:nuc_details}a). A H4 tail
on one nucleosome may interact with an H2A/H2B acidic patch on a adjacent
nucleosome, acting as a tether connecting the two nucleosomes
\cite{Kan2009,Kalashnikova2013}. The H2A and H2B tails, much shorter than their
H3 and H4 counterparts and the subject of a much smaller literature, do not seem
to significantly contribute to internucleosome interactions, although they are
required for proper nucleosome reconstitution \cite{Bertin2007}.

\subsubsection{Histone tail post-translational modifications (PTMs)}

\begin{figure*}
  \includegraphics[width=\textwidth]{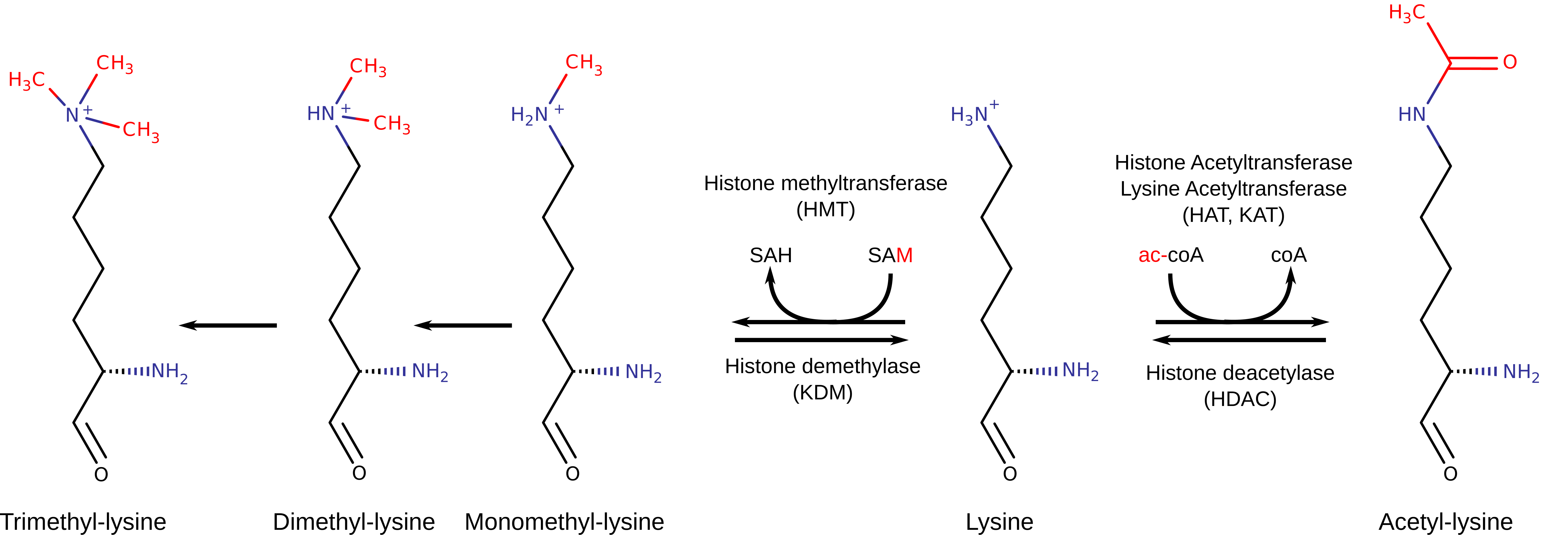}
  \caption{
  Chemical formulas of unmodified, acetylated, and mono, di, tri-methylated
  lysine residues. Methylation (from lysine to the left) is achieved by
  successive additions of single methyl groups, using the SAM metabolite as a
  source of methyl groups. Acetylation (from lysine to the right) is achieved
using the acetyl-coA cofactor as a source of acetyl groups.  Acetylation reduces
the charge at biological pH, whereas methylation preserves the charge.
\label{fig:chemistry-PTM}}
\end{figure*}

Histone tails, besides their role in the structuration of nucleosome arrays, are
also the support of virtually all PTMs targetting histones, which consist in
replacing groups of atoms on one residue by another, chemically different one
(see Fig.~\ref{fig:chemistry-PTM}). For an historical account of their
discovery, see \cite{Morange2013}.  The globular histone core and the lateral
surface of the nucleosome may also undergo post-translational modifications,
which modulate the nucleosome stability, DNA wrapping \cite{Tropberger2013,
Tessarz2014}, hence chromatin architecture. The repertory of histone-tail PTMs
is vast both in terms of types of modifications and in terms of where the
modification can take place \cite{Pepenella2013, Zentner2013, Fierz2014}.

In order to reach a comprehensive physical picture we oversimplify the daunting
complexity of epigenetic histone PTMs \cite{Kouzarides2007} to focus here on:

(i) lysine methylation and notably the two main histone PTMs that are involved
in gene silencing: tri-methylation of the lysine 9 of H3, noted H3K9me3, which
recruits HP1 and tri-methylation of the lysine 27 of H3, noted H3K27me3, which
recruits the Polycomb architectural complex;

(ii) lysine acetylation and specifically the acetylation of lysine 16 of H4,
H4K16ac, which is a hallmark of active chromatin (actively expressed genes).

Epigenetic marks are deposited on or removed from histone tails by dedicated
enzymes, so-called ``writers'' and ``erasers'' \cite{Fierz2014}). Writers
devoted to acetylation are histone acetyltransferases (HAT), notably lysine
acetyltransferases (KAT), and writers devoted to methylation are histone
methyltransferases (HMT), notably lysine methyltransferases (KMT). Erasers are
histone deacetyltransferases (HDAC) and histone demethyltransferases (HDM),
notably lysine demethyltransferases (KDM), see Fig.~\ref{fig:chemistry-PTM}.
A comprehensive list of the known writers and erasers is given in table
\ref{tab:histone_ptm}.

\begin{table*}
\centering
\caption{Principal histone lysine acetylations and methylations of the human
epigenome and their writers/erasers (adapted from \cite{Yang2007a,
Kouzarides2007, Sapountzi2011, Zentner2013, Pepenella2013, Fierz2014}). Other
PTM not shown include arginine methylation and acetylation, serine/threonine
phosphorylation, ubiquitination, and crotonylation (see
\cite{Sadakierska-chudy2014}).} \label{tab:histone_ptm}

\resizebox{\textwidth}{!}{
  \begin{tabular}{ l l l l }
    Type & Family/Class & Proteins/Complexes & Targets \\
    \hline
    \hline
    & GNAT & HAT1 & H4(K5, K12) \\
    & GNAT & PCAF & H3(K9, K14, K18) \\
    Acetyltransferases & GNAT & SAGA & H3 \\
    (HAT/KAT) & CBP/p300 & CBP/P300 & H3(K14, K18), H4(K5, K8), H2AK5, H2B(K12, K15)\\
    & MYST & TIP60 & H4(K5,K8, K12, K16), H3K14 \\
    & MYST & HB01 & H4(K5, K8, K12) \\
    & MYST & MOZ,MORF,MOF & H4 \\
    \hline
    & Class I & HDAC1-3,8 & H3, H4 \\
    & Class IIA & HDAC4,5,7,9 & H3, H4 \\
    Deacetylases & Class IIB & HDAC6,10 & H3, H4 \\
    (HDAC) & Class III (sirtuins) & SIRT1-7 & H3, H4, H4K16 \\
    & Class IV & HDAC11 & H3, H4 \\
    & \textit{none} & SIN3-HDAC1,2 & H3, H4 \\
    & \textit{none} & N-COR/SMRT-HDAC3 & H4 \\ 
    \hline
    & SET & MLL1-5 & H3K4 \\
    & SET & SET1A, SET1B & H3K4 \\
    & SET & G9A/GLP & H3K9 \\
    & SET & SETDB1 & H3K9 \\
    Methyltransferases & SET & SUV39H1, SUV39H2 & H3K9 \\
    (HMT/KMT) & SET & EZH2 (PRC2) & H3K27 \\
    & SET & NSD1 & H3K36 \\
    & SET & SET2 & H3K46 \\
    & SET & SUV420H1, SUV420H2 & H4K20 \\
    & \textit{none} & DOT1L & H3K79 \\
    \hline
    & LSD1 & BHC110 & H3K4 \\
    & LSD1 & COREST-LSD1 & H3K4 \\
    Demethylases & JmjC & JHDM1A, JHDM1B & H3K36 \\
    (HDM/KDM) & JmjC & JHDM2A, JHDM2B & H3K9 \\
    & JmjC & JMJD2A, JHDM3A & H3(K9, K36) \\
    & JmjC & JMJD2B & H3K9 \\
    \hline
  \end{tabular}
}
\end{table*}

A wealth of data exists regarding the presence of histone tail modifications in
different species, development stages and cell types -- the so-called epigenome
-- but efforts for characterizing the effect of histone PTMs are currently
limited by the difficulty of examining {\it in vivo} chromatin structure.
Interestingly, the two main modifications discussed here -- lysine acetylation
and lysine methylation -- seem to act on the chromatin architecture and state of
activity through rather different mechanisms. In the case of  acetylation, a
direct effect on nucleosome-nucleosome interactions is at play, with a certain
but subtle relationship with the associated loss of a positive charge (see
Sec.~\ref{subsec:acetyl}).  In contrast, methylation preserves electric
charges, while introducing significant steric hindrance and potentially
hydrophobic interactions, and mainly act on chromatin indirectly by recruiting
additional architectural proteins (see Sec.~\ref{subsec:methyl}). For this
reasons, acetylation mechanisms are more easily studied by {\it in vitro}
experiments, while methylation effects are more generally studied in the {\it in
vivo} context in presence of their multiple partners. We will now sum up some of
the main experimental results and theoretical interpretations concerning both
these PTMs.

\subsubsection{Histone tail acetylation: direct effects on chromatin accessibility}
\label{subsec:acetyl}

\paragraph{Experiments}

Experimental studies of the role of histone tail acetylation in the architecture
of nucleosomal arrays are conducted using reconstituted, {\it in vitro}
chromatin. In this approach, nucleosomes are reconstituted by incorporating
recombinant histones with tailored aminoacid sequences on tandem repeats of a
DNA sequence with very high histone affinity (the so-called ``601 sequence'').
The sedimentation coefficient of such arrays is then measured as a proxy for
their folding propensity, comparing the sedimentation coefficient of arrays with
of without combinations of histone tail acetylation \cite{Shogren-Knaak2006,
Wang2008, Liu2011, Allahverdi2011}.  In addition, small-angle X-ray scattering
assays on folded nucleosome arrays give estimations of internucleosome
interaction energies \cite{Bertin2007c, Howell2013}. Taken together, these
studies show that H4 tail acetylation decreases internucleosomal intra-array
associations \cite{Hizume2010}.

Acetylation of lysine 16 of histone H4 (H4K16ac) has the strongest effect in
this regard, and may lead to massive disruption of dense chromatin fibers {\it in
vitro} \cite{Shogren-Knaak2006}.  Structural effects of H4K16 acetylation on
chromatin compaction are also confirmed by the observation of a weakening of
chromatin packing {\it in vivo} \cite{Shahbazian2007}, and are in general
associated with actively transcribed genes (e.~g.~, \cite{Taylor2013}).

Surprisingly enough, histone H3 acetylation, which also reduces the charge of
the tails, does not seem to modify the folding propensity of nucleosome arrays
\cite{Wang2008} pointing to a specific mechanism of H4K16 acetylation.

\paragraph{Models}

Experimental studies are often combined with computational models to provide
deeper insights on how the electrostatic nature of histone tail PTMs influence
chromatin folding. 

\begin{figure*}
  \includegraphics[width=\textwidth]{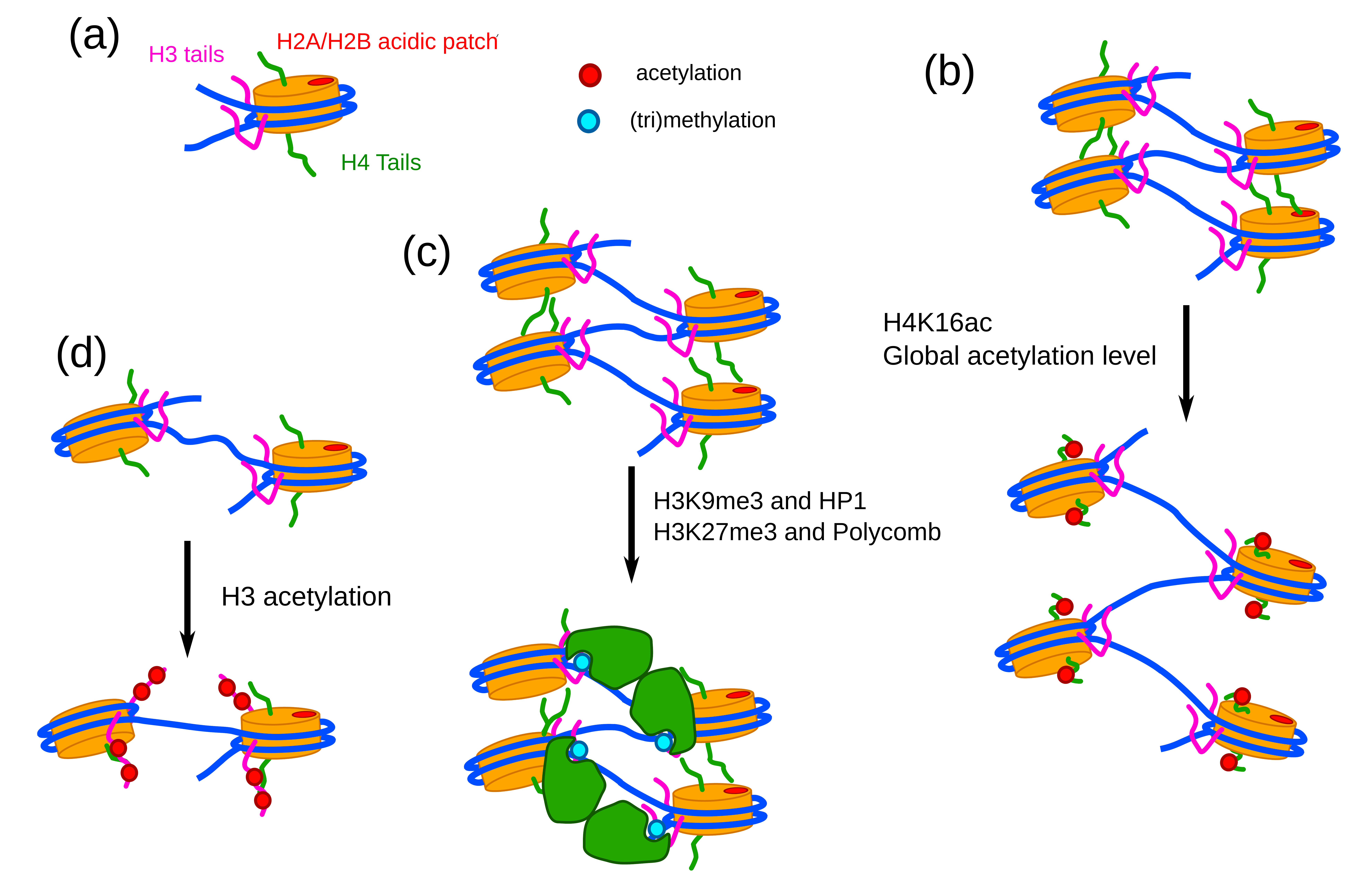}
  \caption{
    Nucleosome arrays and histone PTMs. (a) Cartoon of a nucleosome core
    particle (histone core in yellow, DNA in blue) (b-d) Examples of chromatin
    structural modulation by histone PTMs. Right: H4K16 acetylation induces the
    collapse of the N-terminal H4 tails on their own nucleosomal DNA, preventing
    them from binding the H2A-H2B acidic patches of adjacent nucleosomes.
    Middle: H3 methylation recruits chromatin associated proteins to form
    heterochromatin (e.g. H3K9me3 and HP1, H3K27me3 and the Polycomb family
    complexes). Left: acetylation of H3 tails decreases their affinity for
    nucleosomal or linker DNA and reduces the electrostatic screening of DNA
    negative charges, leading to changes in the mechanical properties of the
linker and accessibility of nucleosomal DNA for other proteins. \label{fig:PTM}}
\end{figure*}

Potoyan \& Papoian \cite{Potoyan2012} addressed the question of the decompaction
induced by H4K16 acetylation, and carried out all-atom simulations in explicit
solvent to compare the conformation of H4 tail with and without this
modification. For the isolated histone tails, H4K16ac leads to slightly more
compact and significantly more structured globular H4 tails. At this level,
compaction is not surprising since the net charge reduction weakens self
repulsion between the tail residues. When DNA is present, i.e. when the entire
nucleosome is considered, tails have a similar behavior: acetylated tails are
more compact, less fluctuating, and are more frequently bound to their own
nucleosomal DNA.  However the less charged acetylated tail interacts much more
strongly ($\sim 5-6 k_BT$) with DNA than the unmodified one ($\sim 2 k_BT$), in
contrast to what is expected from electrostatic reasons. This counterintuitive
effect is achieved thanks to an important tail reorganization that brings other
lysines closer to DNA. While the overall electrostatic attraction is basically
unchanged, the collapse of the tail is favored by hydrophobic interaction and
entropic gain.  In contrast, unmodified H4 tails are more extended and flexible.
They showed a preferential interaction  with linker DNA \cite{Angelov2001} and
with an acidic patch exposed on the surface of next H2A/H2B dimers of
neighboring nucleosomes \cite{Zhou2007} (see Figs.~\ref{fig:nuc_details}a and
\ref{fig:PTM}a).  Hence, while modified H4 tails may contribute to the
nucleosome-nucleosome attractive interaction by the so-called ``tail bridging''
effect \cite{Muhlbacher2006}, the acetylation of lysine 16 might oppose this
effect, leading to weakened nucleosome--nucleosome interactions
\cite{Potoyan2012} (see Fig.~\ref{fig:PTM}b,c,d). Of note this is
qualitatively consistent with experiments on the disordered C-terminal tail of
the p53 protein where a significant increase of its \emph{site-specific} DNA
binding is observed both {\it in vitro} and {\it in vivo} when acetylated
\cite{Luo2004}.

Other computational models generally rely on coarse-grained approximations of
the nucleosome core particle and linker DNA which integrate the mechanical
dynamics of nucleosome as well as its distribution of charges. Arya \& Schlick
used their Discrete Surface Charge Optimization framework to provide estimations
of the contribution of tails to electrostatic interaction energies, showing that
H3 tails principally screen the negative charge of linker DNA, while H4 tails
mediate internucleosomal interactions \cite{Arya2006, Arya2009}, in agreement
with previous experimental findings. However, these studies do not compare
interaction energies with or without histone PTMs. Several other coarse-grained
models have been used so far to specifically investigate histone tail
acetylation \cite{Yang2009,Allahverdi2011,Liu2011}, showing that the effect of
PTMs also largely depend on the valency and the concentration of  bulk
counterions, consistent with sedimentation assays.

\subsubsection{H4K16 deacetylation is a silencing mark in budding yeast}

In budding yeast, and this is specific to budding yeast, silencing is not
achieved by histone methylation. Instead heterochromatin is induced by silent
information regulatory (SIR) complexes which are recruited by deacetylated
nucleosomes. This mechanism is crucially relying on H4K16 \cite{Dayarian2013}
(see below Sec.~\ref{seq:one-dimensional} for the detailed mechanism). 

\subsubsection{Histone tail methylation: indirect effects on chromatin condensation}
\label{subsec:methyl}

In animals, notably in drosophila and mammals, silencing is mainly achieved
through histone tail methylation which, as mentioned above, does not directly
induce chromatin fiber compaction (a notable exception was reported in
\cite{North2014}) but leads to the recruitment of additional architectural
proteins, typically heterochromatin proteins.

Importantly, such architectural proteins are included in the set of proteins
that have been used to define the chromatin  colors in drosophila
\cite{Filion2010}. Precisely, chromatin colors are specific combinations of
epigenetic marks and associated proteins belonging to the following set:
histone-modifying enzymes, proteins that bind specific histone modifications,
general transcription machinery components, nucleosome remodelers, insulator
proteins, heterochromatin proteins, structural components of chromatin, and a
selection of DNA binding factors \cite{Filion2010}. Histone tail methylation
seems therefore to act as a (region specific) substrate to recruit (non
specific) proteins. In turn, these proteins induce different chromatin-chromatin
interactions in different regions, and eventually different chromatin folding
leading in particular to different compaction degrees (see
Sec.~\ref{sec:HiC}). 
 
There are various kinds of heterochromatin in animals (e.g. `black', `blue' and
`green' chromatin in drosophila; even more ``colors'' in mammals). We focus
here on the physical mechanisms that drive the two main silencing processes in
animals, namely the recruitment and spreading of HP1 (Heterochromatin Protein 1)
by the H3K9me3 mark \cite{Hathaway2012,Azzaz2014} and the recruitment of the
Polycomb architectural complex (PcG) by the H3K27me3 mark \cite{Tie2009}. We
moreover discuss the role of these architectural proteins in the physical
process of heterochromatinization.

Unlike acetylation, results obtained {\it in vitro} using reconstituted
chromatin arrays are not directly transferrable to {\it in vivo} contexts for at
least two reasons: (i) lysine methylation has no direct physical effect
(recall that, unlike lysine acetylation, lysine methylation does not change
electric charges), instead, lysine methylation is recognized as a biochemical
tag by dedicated chromatin proteins, either architectural \cite{Zentner2013,
Ong2014, Gosalia2014, Mulligan2015} or remodeling proteins (\cite{Becker2002};
(ii) there is considerable cross-talk among histone tail PTMs
\cite{Li2008,Kouzarides2007,Bannister2011}) which can then form networks
comparable to signaling pathways, eventually resulting in a structural effect.
An example of such a pathway is given by \cite{Wilkins2014} in the context of
budding yeast cell division where phosphorylation of Serine 10 of the H3 tail
induces H4K16 deacetylation, which eventually leads to chromatin compaction.

\paragraph{HP1-mediated heterochromatin}

The family of Heterochromatin Protein 1 (HP1) are fundamental components of
heterochromatin. They are abundant at the centromeres and telomeres (which
correspond roughly, as we have seen, to central and ending regions of the
chromosomes, respectively) in nearly all eukaryotes.

They display high binding affinity for the H3K9me3 mark and are therefore
specifically targeted to nucleosomes harboring this mark. However the spreading
of HP1 along an H3K9me3 epigenetic domain is still a matter of debate. Thus in
the latest special issue of JPCM \cite{Everaers2015}, devoted to the physics of
chromatin, two contrasted models have been proposed: the group of Andrew
Spakowitz \cite{Mulligan2015} claims that bridging interaction between HP1
dimers is critical for HP1 spreading, at odds with the group of Karsten Rippe
\cite{Teif2015} who claims that the binding of one HP1 dimer can stabilize a
stacked nucleosome conformation and facilitate the binding of a second dimer via
an allosteric change of the nucleosome substrate, with no need for a direct
interaction between neighboring HP1 dimers.  It is to be noted that both groups
could reproduce the {\it in vitro} binding curves of the yeast analog of HP1
(Swi6) on mono- and dinucleosomes as well as on arrays of nucleosomes. Moreover
Spakowitz's group claims that HP1 bridging interaction between different
chromatin fibers explains the phase separation of heterochromatin and euchromatin
\cite{Mulligan2015}, whereas Rippe's group evidenced a dependence of the binding
stoechiometry on the NRL (nucleosome repeat length) due to allosteric
cooperativity of binding for nucleosome arrays with long but not with short DNA
linkers, pointing to a facilitated spreading of HP1 on long NRL substrates.

\paragraph{Polycomb-mediated heterochromatin}

Polycomb are a family proteins that mediate transcriptional silencing
\cite{DiCroce2013,Simon2013}.  In drosophila, it was found that two distinct
regulatory complexes (PRC1 and PRC2) are able to silence the \textit{Hox} genes
in a stable and inheritable way \cite{Paro1998,Beuchle2001}. It provides a
mechanism for ``cellular memory'' \cite{Ringrose2004}, that has been speculated
to be alternative to DNA methylation \cite{Bird2002}.

The precise mechanism underlying the heritability of the repressed state of
genes silenced by the Polycomb complexes is still debated. It is known that the
repressive histone mark H3K27me3 (see Sec.~\ref{subsec:methyl}) is recruited by
the PRC2 complex. In turn, H3K27me3 recruits PRC1, which then induces histone
H2AK119 ubiquitination. However, recent studies showed that this relationship
may also work in the opposite sense\cite{Blackledge2014,Cooper2014}. It has also
been suggested that in X chromosome inactivation (see
Sec.~\ref{sec:X-inactivation}), histone ubiquitination, and Polycomb proteins
are mechanistically related to propagate the silenced state
\cite{deNapoles2004}.

A physical modelling of the cross-talk between histone marks and the Polycomb
complexes would be useful and is, to the best of our knowledge, still missing.

\subsection{How epigenetic marks organize the chromosomes in the cell nucleus.
General rules. Physical modeling of epigenome wide studies.}
\label{sec:epigenome}

\subsubsection{Epigenome wide studies}
\label{sec:HiC}

\begin{figure*}
  \includegraphics[width=\textwidth]{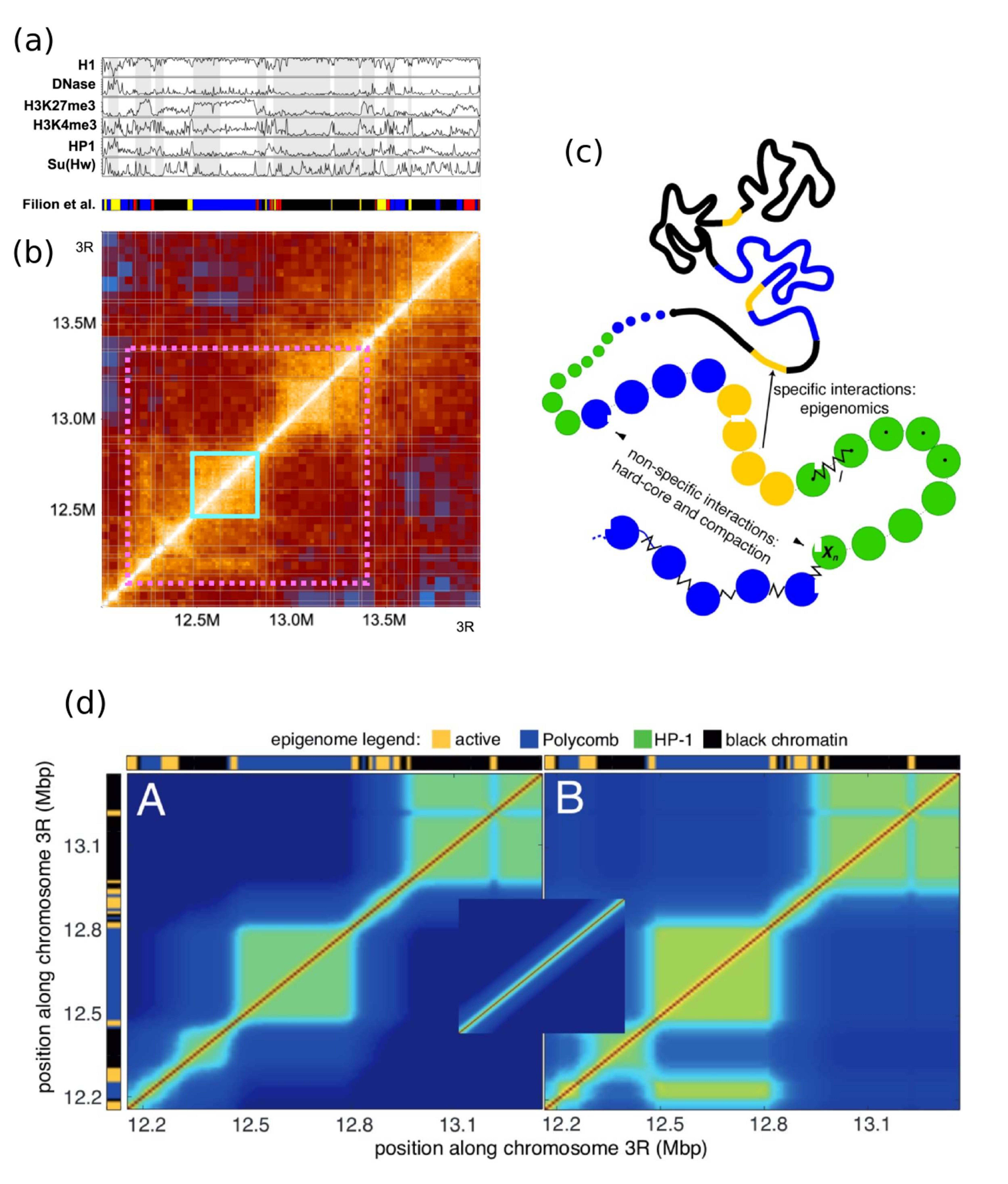}
  \caption{
    Modeling of chromosomal contact maps from the epigenomic landscape.  (a)
    Profiles of H1 occupancy, DNA accessibility, H3K27me3, H3K4me3, HP1 and a
    histone modifier, Su(Hw) along a region of the Dorsophila chromosome 3R.  At
    the bottom, the colors corresponding to these profiles are shown. Yellow and
    red correspond to active chromatin, blue to polycomb bound regions.
    \cite{Filion2010} (b) The corresponding contact map (c) Schematics of
    the co-polymer model used in \cite{Jost2014a} (d) Two predicted contact
    maps corresponding to the region indicated by the pink dashed square in
    (b). ((a), (b) adapted from \cite{Sexton2012}; (c), (d) adapted from
    \cite{Jost2014a})\label{fig:HiC}}
\end{figure*}

One of the current paradigms in the field is that the epigenetic landscape is
driving the 3D genome folding and by extension the functional state of the
cell.  In order to tackle this issue at the genome scale level, epigenomic
techniques based on Next Generation Sequencing (NGS) are increasingly used
\cite{Rivera2013}. These techniques are commonly used to map accessibility,
protein binding sites, and biochemical modification of histones or DNA along
the linear genome (e.g. in drosophila Fig.~\ref{fig:HiC}a). A new technique,
genome-wide Chromosomal Conformation Capture (Hi-C) has been developed in order
to address the issue of genome 3D folding using NGS. This technique allows to
the generation of a list of pair-wise contacts between distal parts of the
genome in various organisms and cell types (e.~g.~ in drosophila,
Fig.~\ref{fig:HiC}b, \cite{Sexton2012}). A high contact probability is
characterized by a bright pixel in the Hi-C contact map (see
Fig.~\ref{fig:HiC}b, around the diagonal). The first results have confirmed the
physical segregation of the genome into heterochromatin and euchromatin regions
\cite{Lieberman2009}. An attractive model of chromosome folding has been
proposed in this seminal paper, named ``fractal globule'', which explains at
the same time the existence of chromosome territories, the absence of knots in
chromosomes, and the power-law decay (with exponent close to $-1$) of the
contact probability as a function of genomic distance \cite{Mirny2011}. This
long-lived metastable state was introduced 20 years before on theoretical
grounds \cite{Grosberg1988} for explaining the kinetics of collapse of
homopolymers. At finer scales, Hi-C further led to the identification of
domains along the genome in which contacts are numerous whereas very few
contacts are established in between different domains. These regions are termed
topologically associating domains, TADs \cite{Dixon2012,Nora2012}. TADs can be
seen as high intensity blocks along the diagonal of the chromosomal contact
maps (Fig.~\ref{fig:HiC}d, cyan squares).  It was then realized that
chromosomes are actually block copolymers, each block corresponding to an
epigenetic domain (compare Fig.\ref{fig:HiC}a and \ref{fig:HiC}b). Combining
Hi-C results with the linear epigenomic annotations of the genome (i.e. the
biological information of the underlying sequences) is in principle a powerful
method to comprehend the functional architecture of the genome. 

Several physical models have been developed so far in order to understand the
3D folding properties of block copolymers. The main goal of these studies is to
recover the chromosomal contact maps observed from the Hi-C data. Two main
classes can be distinguished: simulations that explicitly compute the 3D
chromosome conformations \cite{Barbieri2012,Benedetti2014} and implicit models
in which average contact maps are directly computed in a self-consistent
Gaussian approximation \cite{Jost2014a}. The different explicit models can
account for the formation of TADs, either by preferential binding of co-factors
along specific regions of the genome \cite{Barbieri2012} or by topological
constraints \cite{Benedetti2014} but so far, the direct comparison with
experimental results has only been done using the implicit approach
\cite{Jost2014a}. In this study, the authors use the previously described
colors of chromatin drosophila \cite{Filion2010}, which, as discussed
previously, assign to each subregion of the genome an epigenetic ``color'',
based on the specific protein binding and histone marks found in this region
(Fig.~\ref{fig:HiC}a). They then assign specific pair-potentials (see also
\cite{Saberi2015}) between beads of the same or different colors
(Fig.~\ref{fig:HiC}c) and compute the corresponding contact maps using a
statistical approach previously described \cite{Timoshenko1998}.  With
well-chosen parameters, they were able to retrieve the contacts found
experimentally (Fig.~\ref{fig:HiC}d). An important outcome of their study is
that a fixed epigenetic landscape is compatible with several 3D conformations
of the chain, a phenomenon which they call multistable folding (see below ``the
physics of TADs''). Note that these copolymer models, all at thermal
equilibrium, deviate drastically from the fractal globule hypothesis.

\subsubsection{The physics of TADs: finite-size effects in the coil-globule
transition of copolymers} \label{sec:TAD}

\begin{figure}
  \centering
  \includegraphics[width=0.45\textwidth]{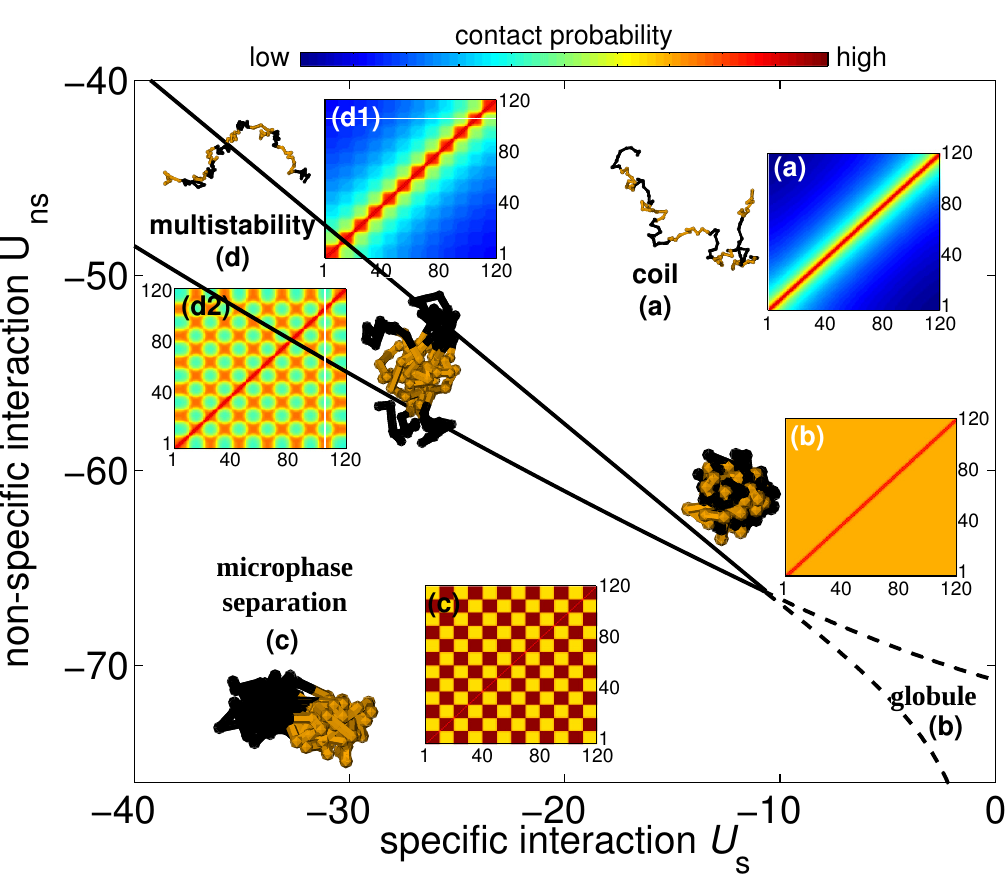}
  \caption{Phase diagram of a toy model copolymer, as a function of specific and
  non-specific interactions. From \cite{Jost2014a}.\label{fig:Jost2A}}
\end{figure}

Jost et al.~\cite{Jost2014a} show a phase diagram of a toy model copolymer as a
function of the intensity of (i) block-specific and (ii) non-specific
interactions, that we show in Fig.~\ref{fig:Jost2A}. On top of the
coil-globule transition of the whole copolymer, there is also coil-globule
transition restricted to each separate block.  Importantly, both coil and
globule phases coexist in a region of the phase diagram, the size of which
depends on the (average) size of the blocks. This is consistent with the
finite-size scaling analysis of the coil-globule transition which has been
proposed in \cite{Care2014} (see also arXiv: arXiv:cond-mat/0004273).

Let us show that both transitions, namely the coil-globule transition inside a
given block and the segregation of different blocks of the same color into
separated microphases, overlap in the phase diagram because of finite-size
effects.

We first remember that a polymer of $N$ monomers, with monomer-monomer attractive
interactions, undergoes a coil-globule transition around the critical
temperature $\Theta (N) = \Theta (1-b \sqrt { \ln (N) / N })$ where $b$ is a
dimensionless prefactor of order unity. More precisely, there is an equilibrium
between coil and globule conformations over a temperature range between $\Theta
(N) - a / \sqrt { \ln (N) }$ and $\Theta (N) + a / \sqrt { \ln (N) }$ where a is
a dimensionless prefactor of order unity. At $T = \Theta (N)$ both coil and
globule conformations are in equal proportions. Therefore, at a given
temperature $T$, longer polymers are more globular than small polymers of the same
kind.

We then consider a copolymer ABAB\ldots made of small blocks A and long blocks
B. For example, the A blocks could be HP1-like heterochromatin, and the B
blocks could be Polycomb-like heterochromatin. We consider monomer-monomer
attractive interactions represented by an energy of interaction $E_{ij}$
between monomers with epigenetic states $i$ and $j$ the following kind: $E_{ij}
= U_{ns} + \delta_{ij} U_s$, where $U_{ns}$ is a non-specific term (does not
depend on $i$ and $j$), $\delta_{ij}$ is the Kronecker delta, and $U_s$ is a
specific interaction term. According to the preceding results on the
coil-globule transition of finite-size polymers, long blocks B go into globules
when small blocks A are still coils. When lowering the temperature (or
equivalently increasing the interactions), blocks B start to transiently bind
together into a macroglobule: this is now the coil-globule transition of the
whole copolymer which is equivalent to a chain of B globules separated by A
linkers; and while this chain collapses (folds) the A linkers start to go into
globules, so that both transitions overlap.

Importantly the macroglobule fluctuates between coil and globule conformations
(as well as any B globule) so that it transiently dissociates thus permitting
the small A blocks, even in remote locations on the genome, to come transiently
into contact (see Fig.~\ref{fig:HiC}). This corresponds to the ``multistate
folding'' region calculated by Jost et al.~and depicted on
Fig.~\ref{fig:Jost2A}.  Note that the width of this multistate folding region
varies as $1 / \ln (n)$ where $n$ is the typical size of the small(est) blocks.

Below the lower critical temperature $\Theta (N) - a / \sqrt { \ln (N) }$ all
the B globules are permanently collapsed in a macroglobule with the A blocks
located at the macroglobule surface (because of interfacial tension). Crucially
the A blocks are still coils, hence in the euchromatin phase, so that their
genomic sequence is expressed, whereas the B blocks are globular and as such in
the heterochromatin phase, hence their underlying sequence is repressed
\cite{Care2015}.

\section{Physical mechanisms involved in the initiation, spreading, maintenance
and heritability of epigenetic marks}
\label{sec:regul-OF}

Stem cells are capable of differentiating to the desired fate depending on the
tissue. Dramatic changes in gene expression occur during development. These
changes are then stabilized and become heritable. Epigenetic modifications take
part in both initiating, stabilizing and propagating the patterns of gene
expression. Gene regulation by epigenetic modifications is indeed stably
propagated through cell divisions (and, in some cases, across generations).  At
each cell division, the whole DNA is replicated. Chromosomes then consist of two
sister chromatids which both have identical genetic information, joined together
at their centromere.  Then, during mitosis, the two chromatids are separated and
segregated into the two nuclei of the daughter cells.

Eukaryotic replication involves both DNA synthesis and chromatin assembly. As
the two double helices are synthesized from the two single strands of the
mother-cell DNA, nucleosomes on the mother-cell DNA strand should also be
distributed to both daughter double helices, and completed by {\it de novo}
nucleosome assembly. In order  to ensure the transmission of epigenetic marks to
daughter cells, mother-cell nucleosomes should be shared by both newly formed
chromosomes, even if the detailed mechanisms of this distribution are still
debated \cite{MacAlpine2013}. 

While it is clear that histone modifications are involved in gene silencing, hence
gene regulation, the questions of how epigenetic marking is initiated, how it may
spread over specific chromosome regions (and not beyond), and how it can be
stably maintained  along the cell cycle and through the cell division are still
under investigation. In this section, we will review the main modeling efforts
that have been made in order to address these questions.

\subsection{Mathematical modelling}
Many recent theoretical works addressed the question of how epigenetic marks are
initiated, spread, and maintained. The main objective of these models is to
reproduce a few essential features observed {\it in vivo}: (a) the
\emph{multistability} of the epigenetic marks; (b) their \emph{spatial patterns}
and (c) their \emph{heritability}.

By multistability, it is generally meant that the epigenetic marks act as
switches between different functional states. In the simplest case, different
patterns of epigenetic marks allow to switch between two states that have a
well-defined functional characterization (bistability).  Such functional states
are then inherited by the daughter cells through mitosis, which is what we call
heritability. As observed in genome-wide studies, the epigenetic patterns
correspond to distinct epigenomic domains that are separated by boundaries (see
Sec.~\ref{sec:epigenome}).

We consider a system of $N$ nucleosomes that can be in $n_S$ different states. In the
simplest case, $n_S = 2$ and one refers to ``modified'' or ``unmodified'' states,
which can be related to active or inactive genes.

The state of the system is described by the variables $\{s_1, s_2, \dots,
s_N\}$, where $s_i$ is the state of nucleosome $i$. If we define $n_j$ as the
number of nucleosomes in the state $j$, then one can write the conservation of
the number of nucleosomes as
\begin{equation}
  \sum_{j=1}^{n_S} n_j = N
  \label{eq:conservation_N}
\end{equation}

Many theoretical works use the silenced mating-type locus of the fission yeast
\textit{Schizosaccharomyces pombe} (reviewed in  \cite{Grewal2002}) as a
model system. In this system, the region containing the two mating-type
regions is normally ``silenced'', i.e. not expressed. The expression of the
mating-type genes may become bistable in mutants, flipping between a silenced
state and an active state  \cite{Grewal1996,Thon1997}. Each
state is stable and heritable; transition between them occurs apparently
stochastically. The \textit{S. pombe} HMT, HDAC  and other proteins are
necessary for silencing, and all may bound H3K9me directly or indirectly.

In the following, we review the models of this behavior proposed so far.

\subsection{Zero-dimensional models}
In zero-dimensional models, neither the spatial organization of the $N$
nucleosomes, nor the notion of distance are introduced. In general, the model
concerns rate equations on how the variables $n_j$ vary as a function of time,
and the objective of the models is to show how bistable or multistable states
can appear. In this class of models, the initiation of the epigenetic mark is
implicitly defined as the initial state of the dynamical system, and the
spreading is described as the time evolution of the initial state. Mitosis
can be modeled as an instantaneous process in which the concentrations of all
species (modified and unmodified nucleosomes) are diluted and the system
restarts. The dilution is due to sharing of mother-cell nucleosomes between both
daugther cells. Nucleosomes are not necessarily shared into equal parts between
daughter chromosomes, but this may be assumed without loss of generality as is
done for convenience in most models.

We can write a general expression for the time evolution of the variables $n_j$:
\begin{equation}
  \frac{\de n_j}{\de t} = \sum_{k\neq j}^{n_S} R_{jk}^+ n_k - \sum_{k\neq j}^{n_S} R_{jk}^- n_j +
  \mathrm{noise}
  \label{eq:0D_general_rate}
\end{equation}
Here, $R_{kj}^+$ is the rate of transition of nucleosomes from the state $k$ to
the state $j$, while $R_{jk}^-$ is the rate of transition from state $j$
to $k$ (obviously, $R_{kj}^+=-R_{jk}^-$). In general, these coefficients are not
constant, but depend on the other dynamical variables. The ``noise'' may be
included to describe the effect of stochastic processes involved in the system.

The simplest possible model of this kind was proposed by
 \cite{Micheelsen2010}. The authors consider the case of $n_S=2$, that is,
they consider only a modified (M) or unmodified (U) state. Using equation
\eqref{eq:conservation_N}, the system may be described by only one variable
$n_M$, the number of modified nucleosomes. The transition rates
are given by
\begin{align}
  R_{UM}^+ & = \alpha n_M^2 + (1-\alpha) \nonumber \\
  R_{MU}^- & = \alpha (1-n_M)^2 + (1-\alpha).
  \label{eq:Micheelsen2010_rates}
\end{align}
This model supposes that the creation of a modified state
involves a cooperative transition (as evidenced by the quadratic terms in
equation \eqref{eq:Micheelsen2010_rates}) or a spontaneous conversion to the
unmodified state (which is described by the $(1-\alpha)$ term). Despite its
simplicity, the model can account for the emergence of bistability. The
parameter $F=\alpha/(1-\alpha)$ (feedback to noise ratio) governs the behavior
of the system. For $F>4$, three fixed points emerge in the system: $n_{M1} = 0$
and $n_{M3} = N$, which are stable, and $n_{M2} = N/2$, which is unstable. The
$F$ parameter is possibly under active control by the cell, which then can
regulate its function (notably by HDAC inhibitors \cite{Dayarian2013}).
Heritability can be partially accounted for by this model, since one can
speculate that cell division brings the system close to the unstable point,
which then returns to its stable attractor.

David-Rus et al.~thoroughly investigated a more general model that still has
$n_S=2$ \cite{David-Rus2009}. 
Their rates read:
\begin{align}
  R_{UM}^+ & = \chi + \alpha n_M^H \nonumber \\
  R_{MU}^- & = \gamma + \eta (1-n_M)^K
  \label{eq:David-Rus2009_rates}
\end{align}
The first interesting result they obtained is that this model can reproduce
bistability only for $H,K > 1$. The simple quadratic case $H = K = 2$ is
a generalisation of the model of Micheelsen et al.~\cite{Micheelsen2010}, where
the cooperative transition probability (rate) from U to M is independent from
that to M to U. If the basal rates $\chi$ and $\gamma$ are small, one again
obtains three fixed points, with the intermediate unstable point being $n_{M2}
\approx \eta/(\alpha+\eta)$. Assuming now that cell divisions exactly halves the
concentration of modified nucleosomes for each daugher cell, if
$n_{M2}>n_{M3}/2$, than the system will always fall in the basin of attraction
of $n_{M1}$ after a cell division, hence the only stable point is the unmodified
state $n_{M1}\approx0$.  Conversely, for $n_{M2}<n_{M3}/2$ (hence
$\eta<\alpha$), the system will converge to the modified state fixed point
$n_{M3}\approx N$ for initial conditions larger than $2n_{M2}$, and bistability
becomes effective.

This scenario is however modified by the presence of noise in the system. In
fact, if the probability of transition from U to M is larger than the
probability of transition from M to U (that is, $\eta<\alpha$), then the
$n_{M1}$ fixed point is no longer stable. Noise drives the system out of the
$n_M = 0$ state, and brings it to the fully modified $n_M\approx N$ state. This
consideration highlights the importance of asymmetric recruitment rates.

\begin{figure}
  \includegraphics[width=.45\textwidth]{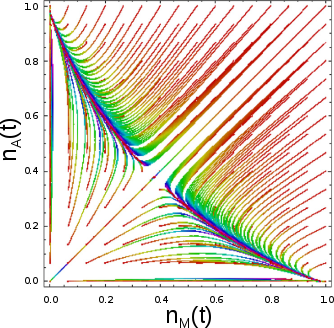}
  \caption{Phase flow portrait for the three-state model described by David-Rus et al.
  \cite{David-Rus2009}, equations \eqref{eq:David-Rus2009_3state_rates}. The values
of the parameters here were $\alpha_A = \alpha_M = 5$, and $\beta_A = \beta_M =
3$. One can clearly observe the presence of the stable fixed points and the
unstable fixed point. From \cite{David-Rus2009}. \label{fig:David-Rus2009-fig4}}
\end{figure}

The same authors also considered the case of $n_S=3$, which was already considered
by Dodd et al.~\cite{Dodd2007} in a very similar form. They consider the case of an
``antimodified'' state (A), that is possibly an acetylated state (active
chromatin mark) that is opposed to the M state which is possibly a methylated
(repressive) state (see Fig.~\ref{fig:models}a). A hypothesis is that only the
$U\to M$ and $U\to A$ are allowed, but the $M\to A$ transition is not
(i.~e.~$R_{MA}^+=0$). They write the following transition rates:
\begin{align}
  R_{UA}^+ & = \alpha_A n_A + \chi_A \nonumber \\
  R_{UM}^+ & = \alpha_M n_M + \chi_M \nonumber \\
  R_{AU}^- & = \beta_M n_M + \gamma_A \nonumber \\
  R_{MU}^- & = \beta_A n_A + \gamma_M. 
  \label{eq:David-Rus2009_3state_rates}
\end{align}
The study of the system in the case where the basal rates $\chi_M$, $\gamma_M$,
$\chi_A$ and $\gamma_A$ vanish already shows the existence of four fixed points:
two stable fixed points, $\{n_A=1, n_M=0\}$ and $\{n_A=0, n_M=1\}$, an unstable saddle
point $\{n_A=0, n_M=0\}$ and an unstable fixed point  $\{n_A= \alpha_M \beta_A / (
\alpha_A \beta_M+(\alpha_M+\beta_M)\beta_A), n_M=\alpha_A \beta_M / ( \alpha_A
\beta_M+(\alpha_M+\beta_M)\beta_A) \}$. The two latter points are aligned along
the $n_A=n_M$ line and create a barrier between the two basins of attraction
\cite{David-Rus2009}. The phase flow diagram of such system is depicted in
Fig.~\ref{fig:David-Rus2009-fig4}.

The last model of this class that we consider is the one proposed by
Jost \cite{Jost2014b}. The author considers a special case of the three-state
model outlined above:
\begin{align}
  R_{UA}^+ & = \epsilon_A n_A + k_0\nonumber \\
  R_{UM}^+ & = \epsilon_M n_M + k_0 \nonumber \\
  R_{AU}^- & = \epsilon_M n_M + k_0 \nonumber \\
  R_{MU}^- & = \epsilon_A n_A + k_0\,, 
  \label{eq:Jost2014b_rates}
\end{align}
that is, it is the same model with $\alpha_{A,M}=\beta_{A,M}=\epsilon_{A,M}$ and
$\chi_{A,M}=\gamma_{A,M}=k_0$. Interestingly, this particular choice allows to
map the system to the zero-dimensional Ising model, with, e.g., the
correspondence A$=+1$, U$=0$, M$=-1$. Within this analogy, recruitment
corresponds to coupling between spins and random transitions are associated with
thermal fluctuations.  A new observable, equivalent to the magnetizaton in the
Ising model, is introduced here: $\mu = a-m$. 

Some known results can thus be recalled for the symmetric recruitment case
$\epsilon_A =\epsilon_M=\epsilon$.  Similarly to what previously discussed,
three fixed points exist. The first one, $\mu = 0$, is stable for weak
recruitment, i.e. for $\epsilon < \epsilon_c = 3 k_0$. Above this critical value
of $\epsilon$, $\mu = 0$ becomes unstable and bistability settles down with the
appearence of two stable fixed points, $\mu_\pm = \pm
(k_0/\epsilon)\sqrt{(\epsilon/k_0 + 1)(\epsilon/k_0 -3)}$.

The non-local character of the nucleosome-nucleosome interaction, which is the
main hypothesis of the zero-order models, has been further justified by a recent
work \cite{Zhang2014b}. The authors proposed a two-layer Potts model in which
in one layer they describe the nucleosomes, and in the other they include
explicitly the enzymes that modify the nucleosomes. The interaction between the
nucleosomes is the effectively mediated by the modifying enzymes. Interestingly,
by integrating out the effect of the modifiers, it is possible to prove the
exact equivalence to the model proposed by Dodd et al.~\cite{Dodd2007}.

To conclude this section, we stress the main results of this comparative
analysis. Bistability is obtained by this class of models in two ways: in
two-state models only when including nonlinear rates, and in three-state models
even having linear rates. The reason for this is that in three-state models the
transitions from a modified to an antimodified state can proceed only in a
two-step process, effectively requiring cooperativity, hence producing
bistable states  \cite{Dodd2007}.

\subsection{Higher-dimensional models}
An inherent limit of the models discussed above is that they cannot reproduce
spatial patterning of the epigenetic marks. Hence, limitation of the mark
spreading should be included by limiting the extent of the concerned domain,
i.e. the total number of nucleosomes. If this assumption may be relevant e.g.
for the mating-type loci in yeast, it probably fails when multicellular
organisms are considered. It is known, for example, that nearly all
noncentromeric H3K9me3 domains in mouse embryonic stem cells have a peaked
shape, with continuously decaying mark densities on both sides
\cite{Hathaway2012}. 

\subsubsection{One-dimensional models.} \label{seq:one-dimensional}
Even when mark spreading is surrounded by boundaries, the question arises how to
model their presence and effects.  Dodd and Sneppen realize in their 2011 work
\cite{Dodd2011} that positive feedback can lead to spreading of the
modifications to genome regions other than the target. They refer in particular
to the silent mating-type loci in budding yeast \textit{Saccharomyces
cerevisiae}. In this organism, the MAT (mating type) gene has two variants, MATa
and MAT$\alpha$, and switching of the mating type occurs when the expressed MAT
variant changes from one type to the other. This mechanism is possible because
each of the two variants comes also with a silenced allele: the HML (hidden MAT
left) carries a silenced MAT$\alpha$ allele, and HMR (hidden MAT right) bears a
silenced MATa allele. The HML and HMR are able to spontaneously flip between
high and low expression states \cite{Xu2006}, thereby allowing for switching of
mating type. These domains are stable over up to 80 cell generations, and are
surrounded by boundary elements that prevent silencing to spread out of the
domains.  These ``barriers'' are specific sequences, and may simply be target
sites for certain DNA-binding proteins, strong gene promoters, or
nucleosome-excluding structures.  Dodd an Sneppen therefore consider a model in
which all nucleosomes are explicitly treated, and the long-distance interaction
between nucleosomes is modeled in a ``local-local'', ``local-global'', or
``global-global'' scheme (see Fig.  \ref{fig:models}c). To limit the long-range
interaction between DNA sites one can introduce a distance dependent
cooperativity, i.e. by making the reaction rate $R_{UM}$ dependent on the
nucleosome distance. A power law dependence, typical of the three-dimensional
probability of contact, can be assumed.

Then, the confinement of silenced regions can be obtained by introducing local
barriers, modeled as  single nucleosomes fixed in the active (A) state. Due to
the local character of the modification step, a single silencing-resistant
nucleosome (e.g. H3K4me3 \cite{Venkatasubrahmanyam2007}) or a
nucleosome-depleted region (notably in gene promoters \cite{Bi2004}) is
enough to stop the silencing spreading, provided that the flanking regions are
entirely in the active state. However, an occasional inactivation of the barrier
make the silencing spread out. This effect can be limited by introducing
regularly spaced weak barriers, modeled as anti-silencers (enhancers) of the U
$\to$ A reaction, or by implementing in the model a Michaelis-Menten saturation
effect when the number of U state nucleosomes increases. The combination of both
effects results in robust prevention of silencing spreading.  

Focusing instead on mammal silenced regions, \citet{Hathaway2012} were able to
reproduce the sharp peaks observed in the experimental modification patterns by
including a ``source'' term in their model. This is a model
in which the initiation and spreading are explicitly separated, and in turns
this allows to reproduce spatial patterning. They write rates as
\begin{align}
U_0  &\xrightarrow[]{k_+^{\mathrm{target}}} M_0 \nonumber \\
\{M_i,  U_{(i+1 \mathrm{~or~} i-1)}  \}  &\xrightarrow[]{k_+} \{M_i,  M_{(i+1
\mathrm{~or~} i-1)}\} \nonumber \\
M_i  &\xrightarrow[]{k_-} U_i.
\end{align}
This description means that at site 0 there is an active modification source
with rate $k_+^\mathrm{target}$, which then spreads to the neighboring
nucleosomes with rate $k_+$. Fitting to experimental results leads to $k_+$ and
$k_-$ rates both of the order of 0.1--0.2 h$^{-1}$ (in agreement with different
experimental estimates of $k_-$. However, as pointed out by the authors, this
model fails to predict the bistable nature of the system, thus not allowing to
describe this crucial feature.

In Ref.~\cite{Hodges2012} a more detailed study of the model is presented.  The
source term ensures that the resulting mark distribution are peaked at the
nucleation site, as experimentally observed, provided
$k_+^{\mathrm{target}}/k_-$ is large enough ($\gtrsim 0.2$), with increased
amplitude and formation rate for increasing $k_+^{\mathrm{target}}$.

Still referring to the mating-type loci in budding yeast, \citet{Dayarian2013}
consider a four-state model with site-dependent rate equations. The fourth state
they consider is a double-acetlyated state, which would correspond to
acetylation of two H4K16 sites.  Importantly, in this model, the modified state
M is supposed to be a state where nucleosomes are bound to silencing (Sir)
proteins, and depends therefore on their availability (concentration). In its
most general form, this is a one-dimensional model that explicitly describes
cooperative transitions that involve any nucleosome pair.  However, it can be
simplified into a zero-dimensional model when considering uniform solutions,
which again show bistability and a characteristic bifurcation diagram. Moreover,
such concentration-dependent model allows for additional interesting effects,
involving a fine balance between the silencing of mating-type loci, which have a
definite extent, and of the telomeres, whose extent may vary depending on the
protein availability \cite{Dayarian2013}.  Interestingly, indeed, this model
also allow the existence of a silenced and an active domains in stable
coexistence, i.e. with an immobile boundary domain, whose position depends on
the balance between environmental self-adjusting parameters as the concentration
of active proteins, a mechanism that these authors explored extensively
\cite{Sedighi2007, Mukhopadhyay2013, Dayarian2013}.

\subsubsection{Three-dimensional models.}
Erdel et al.~\cite{Erdel2013} addressed some more specific questions about the
establishing of epigenetic domains, as how are the chromatin modifying enzymes
targeted or excluded from given chromatin regions, and how exactly the
modification can propagate from one nucleosome to another, how is this state
reestablished or maintained during replication. The proposed model focuses on
the permanent binding of enzymes to a scaffold, either on chromatin itself or on
the nuclear membrane, this leading to the definition of a limited chromatin
region allowed to interact with the enzyme by short-range diffusion. The spatial
distribution of the enzyme hence may result in a spatially limited enzymatic
activity, and results in the definition of epigenetic domains. This first
attempt to take into account the chromatin architecture in a three-dimensional
model is noteworthy, despite the difficulty in estimating many of the
geometrical and physical parameters involved in the model, as the linear
base-pair density along the chromatin fiber, the fiber stiffness, or the
nucleosome local density.  Moreover, the question of how the set up of the
correct architecture in the initial enzyme binding and in defining the
functional chromatin domains remains open.

It also has been recently proposed that pericentromeric heterochromatin spreads
its silenced state with a ``nucleation and looping'' mechanism
\cite{Muller-Ott2014}. Chromatin-bound SUV39H1/2 complexes would act as
nucleation sites and propagate a spatially confined heterochromatin domain with
elevated H3K9me3 modifications via chromatin loops. It is therefore relevant to
include three-dimensional structure in the theoretical modeling of the
spreading of epigenetic marks.

\subsection{Biological relevance of the models} \label{seq:biological-relevance}
In this section we intend to examine the biological relevance of a few key
points that emerged in the discussion of models of initiation and spreading of
epigenetic modifications.

\begin{figure*}
  \includegraphics[width=\textwidth]{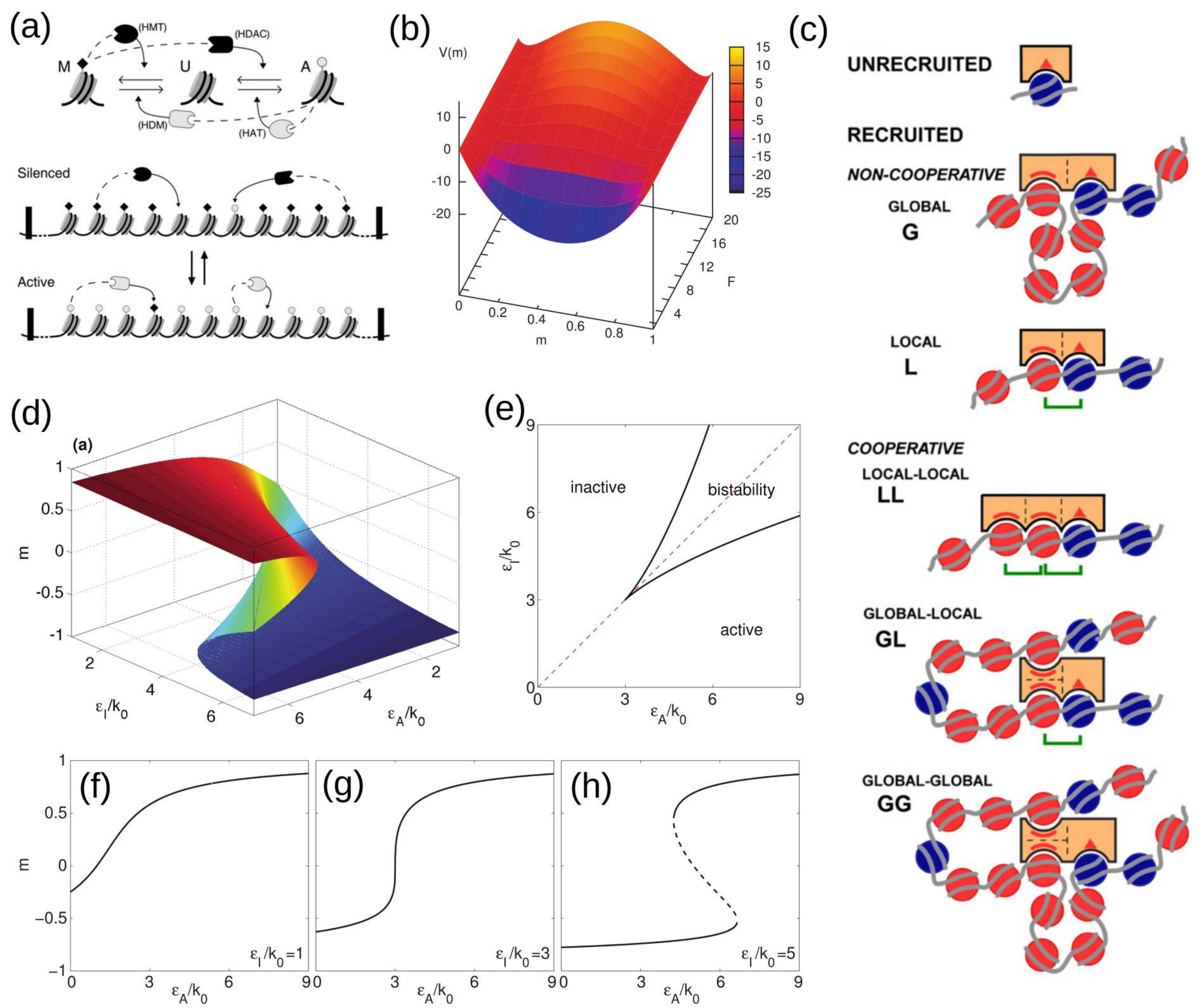}
  \caption{
    A modern view on epigenetic landscapes. (a) The basic mechanism of a
    three-state nucleosome modification model, depicting the modified (M),
    unmodified (U) and antimodified states (A). The transition between M and U
    is catalyzed by histone methyltransferases (HMTs) and histone demethylases
    (HDMs), which depends on an antimodified histone; between U and A the
    transition is catalyzed by histone acetylases (HATs) and histone
    deacetylases (HDACs), which depends on a modified histone. From
    \cite{Dodd2007}.  (b) The coarse-grained potential $V(m)$ as defined in the
    main text, as a function of the mean fractional number of modified
    nucleosomes $m$, and the feedback-to-noise ratio $F$ (from
    \cite{Micheelsen2010}). (c) Different modes of coupling between histone
    modification states, as described in \cite{Dodd2011}. Unrecruited enzymes
    may modify histones directly. Otherwise, recruited enzymes may operate in
    non-cooperative (global or local) or cooperative (local-local, global-local,
    or global-local) modes. From \cite{Dodd2011}.  (d-h)
    Illustration of the ``epigenetic landscape'' as proposed by
    Jost \cite{Jost2014b}. As a function of the control parameters $\epsilon_A$
    and $\epsilon_I$, the system may undergo a transition between an inactive,
    active or bistable state. As shown in (h), the system may also develop a
    hysteretic behavior. (d-h) from \cite{Jost2014b}.
  \label{fig:models}}
\end{figure*}

\subsubsection{Waddington's epigenetic landscape revisited.} First, let us return to
the discussion on the Waddington landscape that we started in the Introduction.
The classical image in the original Waddington representation
\cite{Waddington1957} of a marble rolling down a hill does rather suggest a
fixed landscape, leading to erroneous interpretation when one goes beyond the
metaphorical level (see Fig.~\ref{fig:Waddington}).

In the simplest model we discussed, the one by Micheelsen et al.~
\cite{Micheelsen2010}, the authors show that the model can be reformulated by a
Fokker-Planck equation for the 1D diffusion of a particle in an effective
potential $V(n_M)$ (see Fig.~\ref{fig:models}b). The latter accounts at a time
for drift (external forces) and noise events (with a term of the type $D/\mu$,
with $D$ the diffusion coefficient and $\mu$ the mobility).   
The Waddington idea of an epigenetic landscape is translated in
\cite{Micheelsen2010} in more modern terms, by defining a physically consistent
energy profile. Note however than the mechanism invoked here is not an evolution
along the profile of Fig.~\ref{fig:models}b toward the minimum energy states,
since different values of the $F$ parameter correspond to different system
parameters, hence different external constraints.  In other words, the
equivalent of an epigenetic landscape corresponds here to a given $F =
\mathrm{constant}$ section of the two dimensional potential surface of Fig.
\ref{fig:models}b. This allows in turn to suppose that external constraints may
be included in the parameter $F$, which may vary as a function of metabolism
(level of activity) or drug delivering of ``writers'' or ``erasers'' (see
Sec.~\ref{sec:histonePTMs}), notably HDAC inhibitors \cite{Dayarian2013}, thus
typically making the system switch from bistable to monostable conditions.  As
discussed by Jost  \cite{Jost2014b}, this may also represent a strategy to gain
in system sensitivity hence plasticity during development. Note that the
switching mechanism between bistable and monostable conditions can be
interpreted as the result of an active process bringing the system out of
bistability and favoring its switching to a different state. 

We then stress that it is important to consider the asymmetry of the
modification rates. Taking the notation of Ref.~\cite{Jost2014b} (equations
\eqref{eq:Jost2014b_rates}), we notice that if recruitment of enzymes by
modified or anti-modified marks are different, the stability diagram and the
boundaries between the mono- and bistable regions can be traced as a function
of the two parameters $\epsilon_A$ and $\epsilon_M$. Bistability is observed
only for strong recruitment ($\epsilon_{A,M} < \epsilon_c$) and small asymmetry. 

The epigenetic landscape may also be viewed as a complex, multi-dimensional
dynamical system in which different cell identities correspond to different
dynamical attractors of the system. In one approach, such landscape is modelled
to be shaped by gene regulatory networks. In a recent study, it has been
proposed that stem cell differentiation may be viewed as the process of
transition between steady-state attractors of genes that induce or repress cell
pluripotency \cite{Zhang2014}. Also, it has been suggested that the cellular
identities are characterized by a mixture of several states, and external
signals may drive the transition from one cell state to another one. By
analyzing existing data sets, \citet{Lang2014} have been able to provide direct
evidence for this, demonstrating that epigenetic landscapes are a very powerful
tool to understand cellular dynamics.

\subsubsection{Hysteresis} For even stronger recruitment, a typical hysteretic
behavior appears that may have important biological consequences. One can expect
indeed that, while for differentiated, stable cells recruitment parameters are
almost symmetric, modifications of the environment might actively induce
asymmetric recruitment.  The increase of one recruitment parameter can thus
bring the system along the metastable branch, then make it abruptly switch to
the alternative state, which will then remain stable even when the recruitment
parameters comes back to their initial values, thanks to the hysteretic shape of
the bifurcation curve.  In Fig.~\ref{fig:models}d, starting for instance from
the low $m$ state and symmetric recruitment, one can increase $\epsilon_I/k_0$
and switch to the upper, high $m$ branch, then come back to
$\epsilon_A=\epsilon_I$ without switching back (see also
Fig.~\ref{fig:models}e-h).

Close to $\epsilon_A \sim \epsilon_M \sim \epsilon_c$, the system becomes
ultra-sensitive to perturbations, and highly unstable. This regime may be
associated to diseases. A pathological increase in the frequency of
replication, for instance, may result in an increase of the random transition
rate $k_0$, which in turn may bring the system close to the critical point and
induce epigenetic instability and misregulation.

However, the existence of a critical region may also represent an advantage.
During development, the ability to switch between two coherent states when
applying a weak asymmetric signal (the developmental signal) may facilitate
developmental transitions. Since the random transition rate $k_0$ may be
increased  by reducing the cell cycle, the system can be brought closer to the
critical region and the switch induced by  the application of a weak asymmetric
signal during a finite period of time \cite{Jost2014b}.

\subsection{Example: plant vernalization}

The 3-state model proposed by Dodd et al.~\cite{Dodd2007} has been successfully
adapted to the description on vernalization, the mechanism allowing plants to
flower after a prolonged cold period. 

\begin{figure}
  \includegraphics[width=0.5\textwidth]{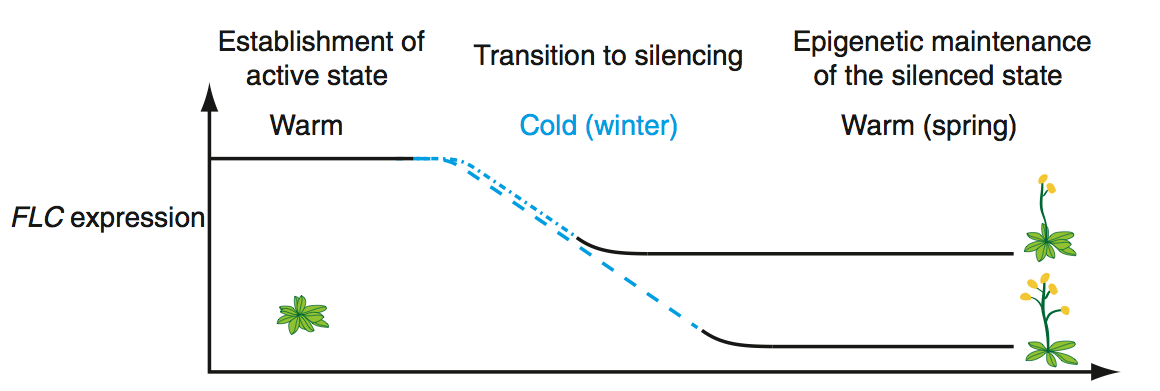}
  \caption{Mechanism of vernalization.  The expression of floral repressor gene,
    FLC, is repressed when plants are exposed to cold and remains stably
    repressed on the return to warm temperatures. Since this repression
    increases with the duration of the exposure to cold, flowering is more
  abundant for longer cold duration (from \cite{J-Song2013}).
\label{fig:vernalization}}
\end{figure}

Plants have the ability to measure the duration of a cold season and to remember
this prior cold exposure in the spring. In \textsl{Arabidopsis thaliana}, an
annual plant, a prolonged cold exposure progressively triggers the
H3K27me3-mediated epigenetic silencing of Flowering Locus C (FLC), a locus
encoding for proteins that in turn act as flowering repressors. The accumulation
of histone epigenetic marks in the FLC locus keeps increasing during the cold.
This slow dynamics of vernalization, taking place over weeks in the cold,
generate a level of stable silencing of FLC in the subsequent warm that depends
quantitatively on the length of the prior cold.  Then, once the FLC is switched
off, the silencing persists at the return of the warm season, and is
mitotically stable through the rest of the development (often for many months)
(see Fig. \ref{fig:vernalization})  \cite{J-Song2013}.  This latter feature is
characteristic of annual plants, while in  FLC perennial plants is repressed only
transiently.

Satake and Iwasa  \cite{Satake2012}  show that this behavior can be accounted
for by means of the Dodd 3-state model  \cite{Dodd2007}, provided that an
explicit dependence on temperature of the model parameters in included.
Explicitly, the transition rates are written in this case as 
\begin{align}
  R_{U \to A} &= \beta \,n_A + \chi  \\
  R_{U \to M} &=u(T) (\beta \,n_M+ \chi)  \\
  R_{A \to U} &= v(T)( \beta n_M + \chi )\\
  R_{M \to U} &= \epsilon (\beta n_A + \chi),
  \label{Satake-rates}
\end{align}
where $u(T)$ and $v(T)$ account for the temperature tuning and takes different
values in warm conditions before vernalization, in cold conditions during
vernalization, and in warm conditions after vernalization. Transition rates are
in fact under the control of a series of proteins (and in particular
Vernalization Insensitive 3, VIN3) whose expression is temperature dependent.
Authors prove that a strong feedback, hence bistability, is necessary to
reproduce the experimental observations.  Interestingly, when the system
evolution is simulated,  the  M $\leftrightarrow$ A transition is observed at a
random time during the cold, for a given system containing  $N$ nucleosomes
(i.e. a given cell).  Different cells switch therefore to the repressed state
after different delays after the change from warm to cold. However, the average
over a cell population leads to a typical behavior that can be reproduced, if
the cell population is large enough \cite{Satake2012}. The duration of winter
memory is also tuned by model parameters, and in particular by those accounting
for to call division rate and rapidity of deposition of epigenetic marks after
vernalization. Changes in these parameters may lead to a much short memory
extent (from more than one year to a few days), this potentially explaining the
different behavior observed in annuals and perennials plants.

While the previous work addressed the question of bistability behavior in
vernalization, the question of the establishment of epigenetic marks induced by
cold is discussed by Angel at al. \cite{Angel2011}, both theoretically and by
experiments  in \textsl{Arabidopsis thaliana}.  These authors focus on the fact
that, when subjected to cold, repression (H3K27me3, M state) only concerns a
small (1 kb) nucleation region inside the FLC (8 kb), close to the first exon
(coding region) after the promoter.  While during the cold only the
nucleation region is marked, after warm restoring the profile changed rather
little in the nucleation region but rose quantitatively across the rest of the
FLC locus according to the length of the cold period.  They ask whether the
small size of the nucleation region  would be sufficient to cause a quantitative
switch in the epigenetic state of the whole FLC locus after return to the warm.
Experimental results are shown to be compatible with a 3-state zero-dimensional
model, provided that two supplementary ingredients are included:  (i) a
site-specific nucleation of the silencing modification during cold, described as
an increased probability to switch to the M state for a sub-ensemble of the
nucleosomes, and (ii) a permanent bias in the histone dynamics towards the M
modification on return to the warm. Within these assumptions simulated
population-averaged levels of the M modification are found to be approximately
stable up to 30 days after the cold, with a modification intensity which depend
on the duration of cold, in good agreement with experiments.

Together, these two studies show that relative simple models displaying a strong
bistability can be usefully employed to model epigenetic mechanisms involved in
real systems as, in the case discussed here, in plants, even if real system
typically includes a few additional features needed to specifically respond to
the particular functional task they are designed for.

\section{Toward a more complex scenario: DNA methylation, role of RNAs,
supercoiling in epigenetics}
\label{sec:more}

Up to now we have focused on histone PTMs and presented them as a crucial issue
in the transmission of epigenetic information. However, the global picture is
more complex. Among the additional epigenetic mechanisms, some are known since
a long time, as DNA methylation (see Sec.~\ref{sec:DNA-methylation}), while others
have been evidenced quite recently, as chromosome coating with (long) non coding
RNAs as in X inactivation (see Sec.~\ref{sec:X-inactivation}), messenger (i.e.
protein-coding) RNA silencing by interaction with micro RNAs (see
Sec.~\ref{sec:microRNA}), or the coupling between epigenetics and supercoiling
(see Sec.~\ref{sec:supercoiling}). An exhaustive description of the overall
picture would represent a titanic task, well beyond the aim of this introductory
review. Therefore we focus here on the main physical aspects of these
biologically relevant mechanisms, drawing on a few concrete examples.

\subsection{DNA methylation}
\label{sec:DNA-methylation}

Historically, DNA methylation has been the first epigenetic mark to be
recognized as a ``stable, inheritable chemical modification that alters gene
expression and does not modify the sequence'' (see Sec.~\ref{sec:intro}). In
fact, in early days of research on DNA methylation, it was found that
methylation states are propagated through mitosis \cite{Wigler1981}.

\begin{figure*}
  \includegraphics[width=\textwidth]{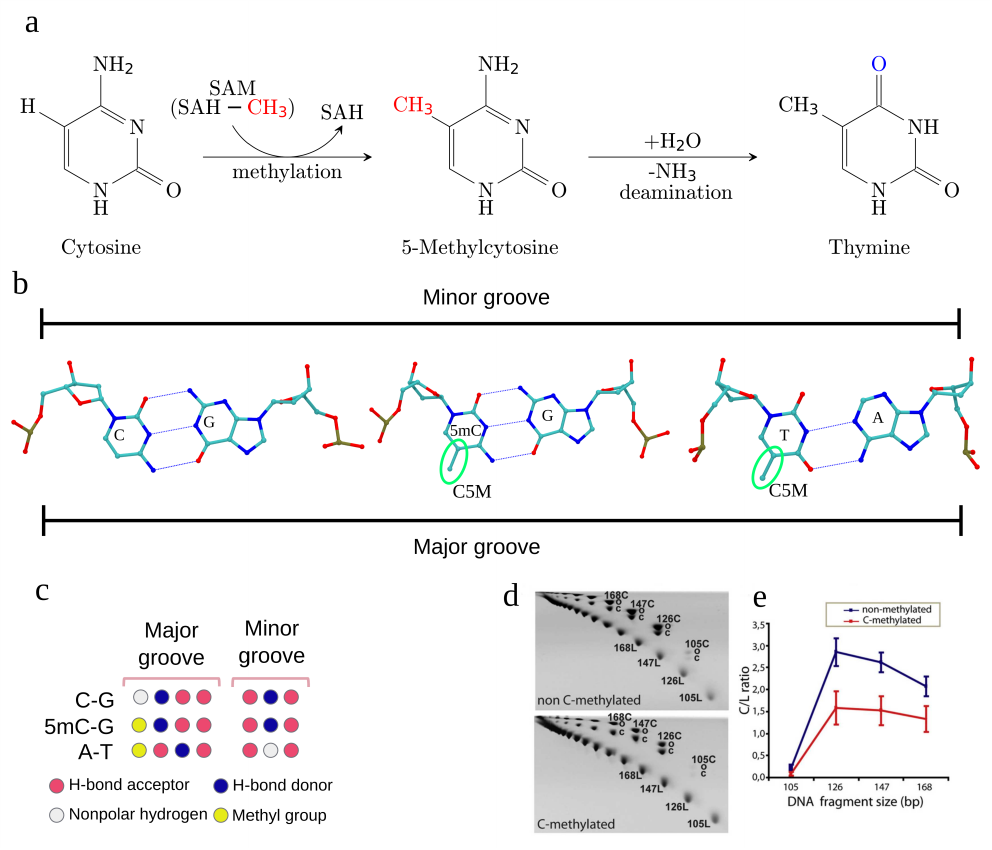}
  \caption{Physical aspects of DNA methylation. (a) Conversion of cytosine to
    5-methylcytosine occurs using SAM (S-adenosylmethionine) as a methyl group
    donor. Spontaneous de-amination may convert 5mC to thymine, leaving the
    methyl group in the major groove. (b) Structural similarities between 5mC
    and thymine in the DNA major groove. (c) Specific patterning of H-bond
    donors, acceptors, methyl groups and non-polar hydrogens allows for ``base
    readout'' of the DNA sequence without strand opening (see main text). (d)
    and (e): experimental cyclization experiments on methylated and unmethylated
    DNA show distinct elastic properties of the two species (figures taken from
    \cite{Perez2012}. In (d), 2D electrophoresis shows the different migrations
    of linear (L) and circular (C) DNA species (either covalently closed (c) or
    nicked open (o)) for nonmethylated and methylated oligomers of 21 bp,
    respectively. (e) Ratio of circular and linear species as a function of
  fragment size. \label{fig:methylation}}
\end{figure*}

DNA methylation is the substitution of a methyl ($-CH_3$) group to the carbon
atom in position 5 at the cytosine base (5mC). Importantly, DNA methylation is
coupled to metabolism through SAM (see Fig. \ref{fig:methylation}a).

The prevalence of DNA methylation in the genome changes significantly among
different organisms: it is very high in vertebrates (where one refers to a
``global'' methylation), very low in \textit{Drosophila}, and absent in the
nematode worm \textit{C. Elegans}.  In somatic cells, cytosine methylation
occurs predominantly at CpG dinucleotides, although it has been detected in any
sequence context both in plants \cite{Cokus2008} and humans \cite{Lister2009},
where 70--80\% of CpG dinucleotides are methylated.

\begin{figure}
  \includegraphics[width=0.48\textwidth]{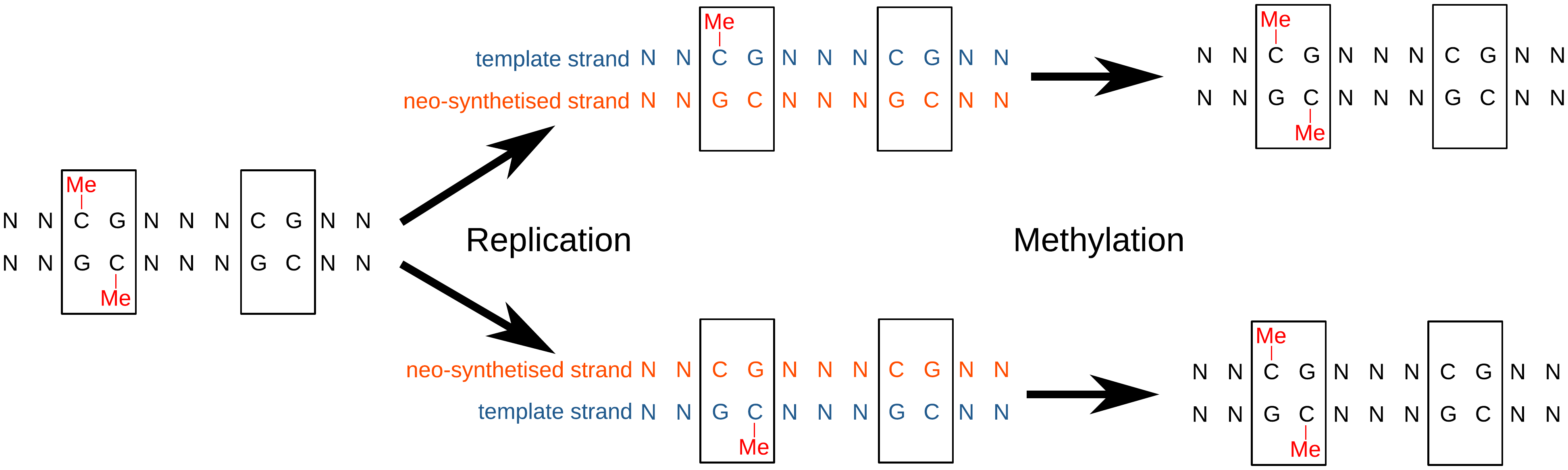}
  \caption{
  Methylation transmission. After replication, only hemi-methylated bases are
  converted into fully methylated bases. \label{fig:methylation-transmission}}
\end{figure}

The patterns of DNA methylation in the genome are established in early
development, and then faithfully propagated throughout successive cell
divisions. Crucially, tissue-specific genes are kept unmethylated, whereas the
others are heavily methylated. These processes are catalyzed by DNA
methyltransferases (DNMTs). It is generally thought that the two
methyltransferases DNMT3A and DNMT3B are responsible for establishing the
methylation pattern during development (\textit{de novo} methylation), and DNMT1
propagates the methylation pattern to daughter cells (\textit{maintenance}
methylation) \cite{Bird2002}, which we illustrate in
Fig.~\ref{fig:methylation-transmission}. The precise mechanism behind the
establishment of the initial methylation pattern during development is largely
unknown. It has been proposed \cite{Khraiwesh2010} that initiation of epigenetic
silencing by DNA methylation depends on the ratio of the miRNAs and their target
messenger RNAs, in a so-called ``RNA-directed DNA methylation''.

5-methylcytosine can convert to thymine by spontaneous de-amination (see Fig.
\ref{fig:methylation}a), leading to a common DNA mutation. The hydrophobic
methyl group in the DNA major groove gives a structural similarity between
thymine and 5mC (see Fig. \ref{fig:methylation}b).  It is important to notice
that this allows for the possibility of a ``base readout'' in the major groove,
as proposed by some authors \cite{Machado2014}.  We shall discuss the
implications of this later.

From the biological point of view, the role of DNA methylation is not clearly
understood. Early studies on the role of DNA methylation highlighted its
importance in gene silencing \cite{McGhee1979,Razin1991}, in X-chromosome
inactivation \cite{Mohandas1981,Graves1982,Venolia1982} and gene imprinting
\cite{Li1993,Razin1994}. It was later established that when CpG-island promoters
are methylated, then the gene will be irreversibly silenced \cite{Jones2012}.
With the advent of technologies that enable for genome-wide screening of the
methylation state of DNA, it has become clear that gene silencing is not the
only role of DNA methylation \cite{Lister2009,Jones2012}, and its biological
role is highly dependent on the sequence context in which it may be found. For
example, DNA methylation has been associated with active gene bodies
\cite{Chodavarapu2010} quite the opposite of its established role of gene
silencing. Detailed analyses of the differences in the methylation patterns in
different cell types revealed an even richer phenomenology
\cite{Lister2009,Spruijt2014,Marchal2014}, suggesting that the role of DNA
methylation is not at all limited to repression of transcription.

DNA methylation has also been implicated in a variety of human diseases (see
\cite{Machado2014} for an extensive review), in particular in cancer, and has
therefore received enormous attention. A challenging issue is the relation of
DNA methylation to cancer progression and prediction of pre-cancerous cell state
\cite{Feinberg2004,Timp2013}.

Can the change in physical properties of DNA upon cytosine methylation help
understanding the variety of its roles? To address this question, we shall
discuss several aspects of the physics involved in cytosine methylation. In the
following, we will review the available knowledge on the following aspects: (a)
the change in the elastic/mechanical properties of DNA upon cytosine
methylation; (b) the role of the hydrophobic methyl group in determining
DNA-protein interactions; (c) the relationship between DNA methylation and
chromatin structure \textit{in vivo}.

\subsubsection{Mechanical properties of DNA change upon methylation}
It has been recently established that the mechanical properties of DNA change
when cytosine is methylated. The extent of this change is still unclear though.

A combination of all-atom molecular dynamics simulations and cyclization
experiments revealed that a single cytosine methylation at a CpG dinucleotide
step has a significant impact on the mechanical properties of DNA
\cite{Perez2012}. Cyclization experiments allowed to prove that oligomers
stiffen significantly upon methylation (see Fig.~\ref{fig:methylation}d-e).
Moreover, the value of the base pair roll was found to increase, whereas the
twist and the width of the minor groove decreased. This should lead to a bending
of the base pair towards the DNA major groove, and a stiffening of the sequence.
However, when a poly-dinucleotide of type d(CpG)$_n$ was methylated, no
significant difference was found compared to its unmethylated counterpart. This
is due to the fact that the GpC dinucleotide has a compensating effect on the
change of mechanical properties of the CpG step.

The stiffening of DNA upon methylation was also predicted by another theoretical
study that employed van der Waals density functional theory \cite{Yusufaly2013}.
There, it was shown that ``Methylation of CG-rich stretches of DNA enhances the
formation of the A-DNA polymorph, a helical form that is more resistant to
bending deformations than B-DNA, and which also bends DNA in the opposite sense.
Consequently, interactions with the histones are inhibited, and nucleosome
formation is suppressed.'' Moreover, by combining single-molecule cyclization
experiments with all-atom molecular dynamics simulations, \citet{Ngo2016}
showed that 5mC increases the DNA local rigidity, and also reached the
conclusion that this destabilizes nucleosomes.

In mixed sequence DNA, no significant effect of DNA methylation was observed by
cyclization of 158--180 bp fragments \cite{Hodges-Garcia1995}. However, when
combining detailed Monte Carlo simulations with cyclization experiments of an
\textit{Eco}RI restriction site, Nathan and Crothers \cite{Nathan2002} found
that methylated sequences change the flexibility and the twist rate of DNA. The
emerging picture is that cytosine methylation changes local structural
parameters of base pair sequences, leaving however unperturbed the global
elastic and mechanical properties of DNA.

Another consequence of cytosine methylation is altered resistance to strand
separation \cite{Severin2011}. By combining single molecule force experiments
with all-atom molecular dynamics simulations, the authors show that strand
separation is strongly affected by cytosine methylation. It inhibits or
facilitates strand separation, depending on the sequence context. Again, the
sequence context plays an essential role in determining the direction and extent
of the impact of DNA methylation.

\subsubsection{Impact of cytosine methylation on DNA-protein interactions}
The addition of a methyl group in the major groove of cytosine bases alters the
hydrophobicity of DNA at the base pair step. In crystallographic studies, it was
found that a methylated A-form DNA oligomer \cite{Mayer1998} is well
hydrated, thereby allowing for the possibility of specific recognition of
methylated DNA sequences, through the interaction with the tightly bound water
at the methyl group.

Three classes of proteins that specifically bind to methylated DNA are known:
MBD (methyl binding domain) proteins, SRA (SET and RING associated-) domain
proteins, and zinc-finger proteins \cite{Buck-Koehntop2013}. Some crystal
structures of proteins that specifically recognize methylated DNA have recently
become available \cite{Buck-Koehntop2012,Liu2014}. The detailed analysis of the
binding modes revealed that although similarities between them exist
\cite{LiuY2013}, it is yet unclear why three distinct families of methylated
DNA-binding proteins were needed in the course of evolution. The sequence
flanking the methylated CpG step was shown to be important in determining the
specificity of the interaction.

The structural differences between methylated and unmethylated DNA may in part
explain the specificity of interactions between proteins and DNA. As shown in
Fig. \ref{fig:methylation}c, the DNA base pair text determines a specific
pattern of chemical groups (hydrophobic, hydrogen bond donor or acceptor) in the
major groove, but not in the minor groove. As a consequence, one proposed
mechanism of recognition of methylated states involves a ``base readout'' in the
major groove, that does not require strand separation \cite{Sasai2010,Zou2011}.

Yet another consequence of DNA methylation is the change in base pair structural
parameters such as twist, roll and minor groove width. \citet{Lazarovici2013}
found that roll and minor groove width were excellent predictors of sequence
specificity for DNaseI endonuclease. This mechanism of recognition of methylated
sequences is termed ``shape readout''. It is likely that both shape and base
readout play a role in most cases \cite{Machado2014}.

\subsubsection{Relationship between nucleosome positioning and DNA methylation}
As discussed earlier, it appears that CpG methylation locally stiffens the DNA,
and it was speculated that as a consequence, nucleosome positioning would be
disfavored. However, contradictory results exist, which also suggest that
methylation actually favors nucleosome formation, or stabilizes DNA wrapped
around a nucleosome. We discuss here the available data.

Several lines of evidence suggest that methylation affects the structure of DNA
wrapped around a histone core. Using fluorescence resonance energy transfer
(FRET) to measure relative displacement between two sites on nucleosomal DNA,
\citet{Lee2011} showed that methylation leads to tightening of DNA around the
nucleosome. Along the same lines, \citet{Choy2010} suggest that the enhanced rigidity
of methylated DNA leads to more compact and closed nucleosomes. Methylated DNA
was also shown to be more difficult to remove from a nucleosome \cite{Kaur2012}.
By combining data from genome-wide nucleosome positioning with available DNA
methylation maps, \citet{Chodavarapu2010} have shown that
there is a small increase (about 1.2\%, from 75 to 76.2\%) in preference of
nucleosome positioning for methylated DNA. This data has been contested by
\citet{Felle2011}, that instead showed that the nucleosome occupancy of
methylated DNA was 2-fold lower compared to unmethylated sequences.

Other studies showed that the increased rigidity of methylated DNA disfavors
nucleosome positioning. We already discussed the works of \citet{Yusufaly2013}
and \citet{Ngo2016}, but many other studies reached the same conclusions
\cite{Felle2011,Perez2012,Portella2013}.

Even more recent studies have shown that methylation has a negligible influence
on nucleosome stability \cite{Langecker2015}.

\subsubsection{Remarks and perspectives}

It has been discovered that 5-methylcytosine is not the only cytosine variant.
In 2009, 5-hydroxymethylcytosine was discovered in mouse brains
\cite{Kriaucionis2009}, and since then, two other forms of cytosine methylation
were discovered: 5-formylcytosine (5fC), and 5-carboxymethylcytosine (5caC) (see
\cite{Spruijt2014} for a review). Each of these other forms has been detected in
mouse embryonic stem cells in significant amounts \cite{C-X-Song2013}, so that 
careful investigation of the differences between their physical properties is
needed, which only very recently has begun \cite{Ngo2016}.

Traditional sequencing methods for detection of cytosine methylation are not
able to distinguish between 5mC, 5hmC, 5fC and 5caC. Accurate methods to
identify the genome-wide map of cytosine methylation states has only very
recently become available \cite{Booth2014}. Much study is needed to understand
the roles of these epigenetic marks.

We speculate that distinctive physical properties of these alternative forms are
in part responsible for determining the variety of roles attributed to DNA
methylation \cite{Spruijt2014,Schubeler2015}. We suggest that physics may help
advancing our understanding of the biological aspects of this important
epigenetic mark. A very recent study has also shown by all-atom molecular
dynamics and single-molecule fluorescence that DNA methylation is able to
promote attraction between DNA fragments mediated by polyvalent cations
\cite{Yoo2016}. This study suggests yet another fascinating role for the role of
physics in determining the large-scale organization of DNA in cell nuclei.

Finally, there is already a vast literature on the coupling between DNA
methylation and histone post-translational modifications (see
Sec.~\ref{sec:histonePTMs}), which was reviewed in \cite{Cedar2009}. However, to
our knowledge there has not been yet an attempt to model this coupling from a
physical point of view. 

\subsection{Parental imprinting}
\label{sec:imprinting}
 
In animals, it has been observed that certain genes are expressed in a
parent-of-origin-specific manner. These so-called ``imprinted'' genes are DNA
methylated on specific sequences named Imprinting-Control Regions (ICRs). The
ICR methylation of any imprinted gene occurs either on the paternally herited or
on the maternally herited chromosome this gene belongs to. Remember that there
is one copy of each gene on the paternal and on the maternal chromosome (except
for sexual chromosomes). Both copies are generally different and are called
paternal and maternal \emph{alleles}. ICR methylation starts in germ cells,
keeps the memory of the parental origin of the allele and drives monoallelic
expression. For example, this mechanism happens on the Igf2/H19 gene locus of
mouse chromosome 7 (see Ref.~\cite{Lesne2015}). The ICR located upstream of the H19
gene is methylated on the paternal allele but it remains unmethylated on the
maternal allele. The maternal unmethylated allele is bound by the
transcriptional repressor CTCF protein (CTCF means CCCTC-binding factor; CCCTC
is a DNA sequence), that prevents the interaction between regulatory sequences
(enhancers) located downstream of H19 and the Igf2 gene located further upstream
of the ICR.  Therefore, the Igf2 gene is not activated on the maternal
chromosome 7. Instead, on the paternal chromosome, the DNA methylation prevents
the binding of the CTCF protein and the Igf2 gene can be activated by the
regulatory sequences. This differential folding has been evidenced by Chromosome
Conformation Capture (3C) experiments during mouse development \cite{Court2011}.

Genomic imprinting was selected during evolution at the transition between
placental (e.g. mouse) and marsupial (e.g. kangaroo) mammals. Six genes in
marsupials and about a hundred in placental mammals undergo parental genomic
imprinting. The reason for the selection of this unusual, epigenetic mechanism
of gene regulation during mammalian evolution remains poorly understood at
present.

\subsection{Chromosome X inactivation}
 \label{sec:X-inactivation}

Another important and historically relevant example of epigenetic silencing is
the inactivation of the X chromosome in mammal females.

The pair of sex chromosomes (XY in males, XX in females) is responsible for sex
determination in mammals. While the Y is small and carries only a few genes, X
chromosome is much longer and contains thousands of genes. Females thus carry
twice as many X-linked genes as males, this leading to a potentially lethal
dosage problem.

During early embryonic development, one of the two X chromosomes of females is
thus inactivated, and condensed to form the so-called Barr body \cite{Barr1949}.

In mice, the inactivation of the X chromosome occurs in three phases. First, the
paternal X chromosome is inactivated during the preimplantation period, from the
stage ``two cells''. Then, it remains inactive in the peripheral cells, which will
form the placenta, but it is reactivated in the cells which will form the future
embryo. Finally, a second inactivation takes place, and this time it concerns,
randomly in each cell, either paternal or maternal X \cite{Okamoto2004}.

Intriguing questions then arise: how does the cell manage to silence only one of
the two X chromosomes? How is the silenced X chosen?

The X inactivation mechanism seems to be controlled by a complex genetic locus
called the X-chromosome-inactivation centre (Xic). It has been proven that
starting mechanism in X inactivation is mediated by the non-coding (not
translated into protein)  transcript of the Xist (X-inactive specific) gene,
present within Xic. Once transcribed, many copies of Xist RNA  accumulate along
the X chromosome (RNA \emph{coating}), then induce its heterochromatinization.

This process is however under the control of a few other genes included in the
Xic region. One of the crucial elements is Tsix, a non-coding RNA gene that is
antisense to the Xist gene (it is transcribed from the complementary DNA
strand). Due to this compementarity, the Tsix RNA-transcript duplexes with the
complementary Xist RNA-transcript into a double stranded RNA which is further
degraded. This mechanism prevents the accumulation of Xist, hence inhibits the X
inactivation  \cite{Okamoto2004}.

Random selection of the inactivated X chromosome may therefore emerge from a
detailed balance in the synthesis of Tsix and Xist.  Recently, a model to
explain this complex regulation path has been proposed \cite{Giorgetti2014},
that relies on the polymer physics properties of chromatin, on its organization
in topologically-associating domains (TADs, see Sec.~\ref{sec:TAD}), and on a
detailed coupling between gene expression and 3D organization at the level of
the Xic center.

The Xic is  composed of two topologically-associating domains (TADs,
\cite{Nora2012}), called the Tsix TAD (320kb) and the Xist TAD (600kb). The 3D
structure of the Tsix TAD is highly variable among cells and this variability is
most probably due to fluctuations of chromatin conformation at time scales
shorter than a cell cycle, as in the model proposed by Jost et al.~previously
discussed \cite{Jost2014a}. The distribution of  conformations observed thanks
to 5C experiments (Carbon Copy Chromosome Conformational Capture), illustrated
by the Fig.~5A in \cite{Giorgetti2014},  shows  indeed an equilibrium between
coil and globule conformations, typical of a  coil-globule transition of a
polymer with finite-size effects \cite{Imbert1997, Care2014}.

The model also assumes that the level of Tsix depends on the activity of two
putative regulatory elements (Linx and Chic1, \cite{Nora2012}), placed inside
the same  Tsix TAD.

Inactivation can then be explained as a result of  the chromatin conformation of
the Tsix TAD, which determines the regulation of specific interactions between
all these elements (regulatory elements, Tsix and Xist). The switch between
globule and coil confirmations changes indeed the spatial proximity between
these genes, hence their interactions.  As a result, globule conformations
induce higher Tsix transcription levels, while coil conformations correspond to
lower Tsix levels.

Statistical fluctuations in chromatin conformation within the Tsix TAD may,
then, contribute to ensuring asymmetric expression from the Xic at the onset of
X chromosome inactivation, as shown by simulations \cite{Giorgetti2014}. If, in
a cell, Tsix TAD is similarly compacted on the two alleles, Tsix transcript
levels from the two alleles are similar, with little or no
heterochromatinization effect.  As a fluctuation induces the coil-globule
transition for the Tsix TAD on one allele, then the two transcripts tend to be
differentially expressed. This mechanism may help ensuring that Xist is only
transcribed from the allele with lower Tsix transcription  (Fig.~6B in
\cite{Giorgetti2014}).

Once established, the X inactivation is stably transmitted through mitosis along
the following development.  Further (and later) features of the inactive X
include hypermethylation of DNA, histone deacetylation, chromatin condensation,
i.e. the same general mechanisms that we have previously introduced.

\subsection{Non-coding RNA and microRNA}
 \label{sec:microRNA}

The X inactivation  is a clear example of the crucial and early role of RNA in
epigenetic silencing. Together with histone and DNA modifications, non coding
RNAs have (more recently) emerged as one of the main epigenetic mechanisms. Two
main classes of epigenetically active non coding RNAs can be identified: small
($<$30 nucleotides) and long  ($>$200 nucleotides). Both classes play a role in
heterochromatin formation, histone modification, DNA methylation targeting, and
gene silencing.

Long RNAs can complex with chromatin-modifying proteins and recruit their
catalytic activity to specific sites in the genome, thereby modifying chromatin
states and influencing gene expression.  In the case of the X inactivation, the
first described epigenetic mechanism involving a long non coding RNA (lncRNA),
it has been shown  that Xist RNA directly recruits chromatin-modifying factors
as Polycomb repressive complex PRC2 that mediates histone H3 lysine 27
methylation, but the direct character of such interaction remains to be
confirmed \cite{Brockdorff2013}, as well as the overall mechanism, including
the interplay between the different RNA involved (Xist Tsix and others)
\cite{Pontier2011}.  According to a proposed model \cite{Lee2012},  long non
coding RNAs may function by sequestering chromatin-modifying enzymes away from
other interacting partners or by  guiding chromatin modifiers to the correct
locations in the genome. In other cases, the long RNA seems to works by binding
and bringing together different types of proteins that can cooperate in
establishing the repressive chromatin state (see \cite{Marchese2014} for a
review on the interplay between lncRNAs and chromatin modifiers in epigenetics).
Very recently, a single molecule, single cell RNA Fluorescence in situ
hybridization (FISH) has allowed to quantify and categorize the subcellular
localization patterns of a representative set of 61 lncRNAs in three different
cell types, giving further quantitative insight in linking their putative role
to their position in the cell \cite{Cabili2015}.  LncRNAs exhibit a diversity of
localization patterns, including a dispersed distribution in the nucleus or in
the cytoplasm, and condensed in nuclear foci, a configuration which may be
consistent with a role for these lncRNAs in chromatin regulation, as shown for
XIST and other lncRNAs involved in imprinting. Most lncRNAs present however
stronger nuclear localization than most messenger RNAs.  Interestingly, lncRNAs
do not persist at nuclear foci during mitosis, suggesting that retention at
specific regulatory regions through mitosis is likely not a mechanism of mitotic
inheritance.

A whole new realm of small non-coding RNAs  was discovered in the late 90s. It
includes two classes of small RNAs: micro RNAs (miRNAs) and small interfering
RNAs (siRNAs), which perform many functions, and in particular are involved in
the so called RNA interference, a regulation pathway of gene expression at the
transcriptional and post-transcriptional level. In other words, RNA
interference may act either by inhibition of the target RNA transcription or by
degradation of the transcript RNA. Piwi-interacting RNAs (piRNAs) represent a
third, large class of lightly longer RNAs, palying a role in epigenetic and
post-transcriptional gene silencing of retrotransposons and other genetic
elements in germ line cells through their association with piwi proteins. Again,
we note here that unnecessary and potentially deleterious genes are often under
epigenetic control \cite{Cowley2010}.

At the post-translational level, the essential repression mechanism for both
miRNAs and siRNAs is through pairing to a complementary sequence of the messenger
RNA transcribed from a target gene, which results in the degradation of the RNA
and thus the repression of the gene. To recognize its target messenger RNA,
miRNAs and siRNAs must be associated with a protein to form the
\emph{RNA-induced silencing complex}.

Interestingly, heterochromatinization may also be initiated by the RNA
interference machinery\footnote{The RNA interference machinery includes
different factors (as Dicer, Argonaute and RNA-dependent RNA polymerases), that
produce the small RNAs or bind them to form functional complexes.} that targets
\emph{repetitive DNA sequences} \cite{Grewal2007}. These are DNA sequences of up
to several million base pairs and consist of a large number of repetitions of a
much smaller sequence, and are found, in particular, at centromeres, telomeres,
and other regions that remain condensed throughout the cell cycle, referred to
as  \emph{constitutive heterochromatin}. At the same time, heterochromatin
mediates the spreading of RNA interference machinery to surrounding sequences,
hence the production of siRNAs, which in turn are essential for the stable
maintenance of heterochromatic structures.

A complex scenario thus emerges in which DNA sequences, RNAs, epigenetic factors
and chromatin remodellers plays together in the setting up and maintenance of
different functional chromatin states.  DNA-methylation and
histone-modifications often act together to regulate miRNA expression, while,
conversely, some miRNAs can regulate the expression of epigenetic machinery,
with important  dysregulation effects in cancer \cite{ZifengWang2013}.

The models previously described  for the spreading of epigenetic marks needs
probably to include these additional features in order to reproduce the
epigenetic mechanisms to a larger scale.  A first step in this direction may be
the study of the interplay between miRNAs and epigenetic regulators, and the
particular role of post-translational regulation, as discussed by Osella et al.~
\cite{Osella2014}.  A typical basic regulatory network involving miRNAs and
epigenetic regulators is the double negative feed-back loop, in which a miRNA
represses an epigenetic regulator, which in turn represses the expression of the
same miRNA.  Starting from an approach similar to what discussed in
Sec.~\ref{sec:regul-OF}, it is possible to describe the system.  More precisely,
the model variables are the number of miRNAs, the number of messanger RNAs that
miRNAs repress, and the number of proteins (epigenetic regulator) resulting from
the messenger RNA translation (and repressing, in turn, the miRNAs).  The
interesting input from the miRNA regulation step is its role in keeping
fluctuations of  gene expression under control, either by suppressing
translation and by promoting RNA degradation. Both effects helps indeed in
reducing the burstiness  in protein production \cite{Friedman2006,Osella2014}.
The resulting set of rate equations includes therefore, on one hand, the highly
non-linear and bistable character of epigenetic regulation and, on the other,
the stabilizing effect of miRNAs regulation, this leading to an increased
stability of the system, and to an increased range of bistability of the switch.

This result suggest possible reasons for the existence of regulatory pathways
combining epigenetic regulators and miRNAs, although both experimental
investigations and modeling of such complicated circuits are still at the
embryonic stage.

\subsection{Supercoilingomics: supercoiling as a physical epigenetic mark, and
its role in the initiation and maintenance of epigenetic marks}
\label{sec:supercoiling}

DNA supercoiling was first properly described by Jerome Vinograd and colleagues
\cite{Vinograd1965} and it took just some more years to James Wang to discover
the first enzyme able to relax these topological constraints in vivo
\cite{Wang1971}. DNA topological state is given by its linking number ($Lk$),
defined as the number of times that a strand of DNA winds in the right-handed
direction around the helix axis when the axis is straight (or constrained to lie
in a plane for a circular molecule). This integer is the sum of two geometrical
parameters: twist (or twisting number $Tw$, a measure of the helical winding of
the DNA strands around each other, hence representing  a ``1D'' deformation
along the axis) and writhe (or writhing number Wr, a measure of the 3D coiling
of the axis of the double helix). The partitioning between $Tw$ and $Wr$ for a
given $Lk$ is determined by the free energy of DNA (itself dependent on ionic
conditions) and by DNA/protein interactions that locally impose some particular
DNA torsion and/or writhe. While structural proteins can only alter the $Tw/Wr$
ratio, enzymes such as topoisomerases or gyrases can alter Lk by catalyzing the
cleavage of one or both strands of DNA, followed by the passage of a segment of
DNA through this break and the resealing of the DNA break \cite{Wang2002}.

In most living organisms, DNA is negatively supercoiled, which prepares DNA for
processes requiring separation of the DNA strands, such as replication or
transcription. In eukaryotes, this negative supercoiling is constrained within
the nucleosome, so that its removal will simultaneously favor the access and
melting of previously occulted DNA, therefore facilitating transcription
initiation. The distribution of nucleosomes, and notably the NRL, appears then
as an important feature to propagate through mitosis, partly for topological
reasons. Regarding the elongation step, as DNA is screwing through the
polymerase during transcription, the negative supercoiling induced in the back
of the enzyme can propagate through the chromatin fiber and trigger local DNA
alterations that have been proposed to serve as a regulatory signal for
molecular partners \cite{Kouzine2004,Liu2006,Kouzine2008,Belotserkovskii2013}.
Therefore supercoiling would act as a transient mechanotransducer as well as a
physical epigenetic mark. Moreover, nucleosome conformational changes might
help to smoothen the elongation process by buffering some topological
constraint \cite{Bancaud2007,Recouvreux2011,Vlijm2015} and facilitate H2A/H2B
dimer loss in front of the polymerase \cite{Sheinin2013}. It remains to be seen
how much the structural differences provided by histone variants
\cite{Shaytan2015} would help to build ``elongation friendly'' regions that
could be transmitted through cell division.

Supercoiling of DNA has been recently proposed to be considered as a true
physical epigenetic mark, entering the family of ``omics'' data one should
consider to get a comprehensive genome-wide epigenetic landscape of a cell at a
given state of its development \cite{Lavelle2014}. Indeed, genome-wide maps of
DNA supercoiling states have been generated
\cite{Bermudez2010,Joshi2010,Kouzine2013a,Kouzine2013b,Naughton2013,Teves2014}
which add to existing predicted maps of DNA melting \cite{Liu2007} or G-quartet
motifs \cite{Du2009,Maizels2013}. The emerging picture is that supercoiling is
associated to the structuration of chromatin topological domains, which largely
overlap with TADs \cite{Naughton2013}.

DNA supercoiling is a physical epigenetic mark because it may change the
affinity of the underlying DNA sequence to specific transcription factors
\cite{Travers2007}. Supercoiling may also silence a whole topological domain
when recruiting TFs which in turn may recruit silencing enzymes, e.g. Suv39h
which eventually deposits H3K9me3 epigenetic marks \cite{Bulut2012}. Note that
in this case, transcription factors are used to repressing instead of activating
gene expression. This mechanism relies on DNA allostery, i.e. the change of DNA
affinity to some transcription factors that is induced by supercoiling
\cite{Lesne2015}.

But how may supercoiling be initiated and herited? The distribution of
topoisomerases and structural proteins such as condensins should help in
transmitting some domain structuration and topological states through cell
division \cite{Aragon2013,Hirano2014}.

The nucleosome repeat length (NRL, see Sec.~\ref{subsec:nucleosome}) might also
be a key control parameter. We first note that the twist rate of the DNA double
helix in a given topological domain is a function of (i) remodeling activity,
notably through active nucleosome removal \cite{Padinhateeri2011}, and (ii)
topoisomerase activity. Importantly both these ATP-consuming mechanisms are
under active control of the cell metabolism. Moreover both remodeling and
topoisomerase activities regulate the value of the average NRL of a given
topological domain.  Therefore the average NRL over some genomic domain appears
to be a physical epigenetic mark of this domain. And the transmission of this
average NRL through mitosis would transmit the twist rate of the domain.
Interestingly the recently observed spreading mode of histone PTMs over
transcription cycles \cite{Terweij2013} might explain the spreading and
maintenance of the NRL on epigenetic domains.  In support of this hypothesis
\emph{active} remodeling processes achieved by ATP-consuming remodeling factors
- and crucially through active nucleosome removal - have been shown to be
essential for driving biologically relevant nucleosome positioning
\cite{Padinhateeri2011}, thus fixing the average NRL. Challenging genome-wide
studies are needed to further correlate supercoiling maps to cell
differentiation states.

\section{Conclusion and perspectives}\label{sec:conclusion}

By putting the rich and diverse  biological literature under the new light  of a
physical approach, the emerging picture is that a limited set of general
physical rules play a key role in the epigenetic regulation of gene expression.
Processes at work  are diffusion-limited and  involve a small number of
molecules, which precludes simple approaches in terms of concentrations.
Instead, multi-scale approaches articulating different models at different levels
of organization are to be developed. 

Mainly, epigenetics  display an intrication of physical mechanisms and  specific
biological entities,  devised in the course of evolution to achieve an
exquisitely coordinated  and adaptable regulation of transcriptional activity.
Our review demonstrates the need to take into account both aspects, within a
dialogue between physics and biology, theory and living-cell experiments.
Importantly, theoretical models are now amenable to experimental verification,
thanks to the many technologies that are available to that end. It is worth
noticing that the field of synthetic chromatin biology \cite{Keung2015}, which
allows to modify and manipulate epigenetic states, is a promising new avenue
for putting theories to the test.

Significant and challenging issues remain:

(i) coupling nuclear architecture and epigenetic marking: understanding the
interplay between 3D nucleus architecture and 1D epigenetic marking (including
barrier positions);

(ii) physical signals that stimulate epigenetic marking: coupling nucleus
mechanical deformations to epigenetic marking. Cells dramatically change their
shape and mechanics during development by integrating physicochemical signals
from the local microenvironment (morphogens gradients, cell-cell contact,
adhesion to extracellular matrix) to generate lineage-specific gene expression
\cite{Engler2006}. Recent studies have begun to uncover the mechanisms by which
these signals are integrated into the 3D spatiotemporal organization and
epigenetic state of the nucleus and impact cell fate decision
\cite{Shivashankar2011,Bellas2014,Ramdas2015}.  Further understanding of these
transduction mechanisms is a challenging perspective;

(iii) equilibrium \emph{vs} nonequilibrium physics: An implicit - yet
overlooked - assumption of the models of bistability introduced by Sneppen and
coworkers in the context of epigenetics \cite{Dodd2007, Micheelsen2010} is that
the system is open and far from equilibrium. The nonequilibrium nature of their
models lies in the asymmetry of the transitions: the recruitment of an unmarked
nucleosome by a marked nucleosome does not affect the latter, in strong
contrast with an equilibrium transition between two species. Indeed epigenetic
marks undergo permanent recycling and biochemical transformations, so that
epigenetic marks turn out to be steady states and not equilibrium states.
Therefore, it is of primary importance to identify, characterize and model the
various active physical mechanisms that are at work in the initiation and
maintenance of epigenetic marks. In particular, it is crucial to evidence
active (ATP-dependent) mechanisms that maintain epigenetic marks, for instance:
metabolism, transcription, replication, ionic pumps at the cell membrane. In
our opinion this is a very challenging and timely topic for biology-oriented
physicists.

\acknowledgments
The authors wish to thank all the members of the CNRS GDR 3536 for the
stimulating discussions that inspired this work. Special thanks to Aur\'elien
Bancaud, Philippe Bertrand, Pascal Carrivain, Giacomo Cavalli, Thierry Forn\'e,
Daniel Jost, and  Cedric Vaillant for their help in the preparation of the
manuscript. This work has been funded by the French Institut National du
Cancer, grant INCa\_5960 and by the French Agence Nationale de la Recherche,
grant ANR-13-BSV5-0010-03.

\bibliographystyle{apsrmp4-1}
\bibliography{RMP-ArXiV-ver3}

\end{document}